\documentclass[12pt]{article}

\usepackage{amsmath}
\usepackage{amsfonts}   
\usepackage{amssymb}    
\usepackage{arxiv}

\usepackage[utf8]{inputenc} 
\usepackage[T1]{fontenc}    
\usepackage{hyperref}       
\usepackage{url}            
\usepackage{amsfonts}       
\usepackage{booktabs}       
\usepackage{nicefrac}       
\usepackage{microtype}      
\usepackage{lipsum}
\usepackage{graphicx}
\graphicspath{ {./images/} }
\usepackage{float}
\usepackage{caption}

\usepackage{listings}
\usepackage{xcolor}

\usepackage{authblk}

\title{Optimal position-building strategies in Competition\thanks{This is version 2 of this paper and contains corrections to typos and improvement in the exposition in certain places.}}

\author[1]{
 Neil A. Chriss
}

\affil[1]{neil.chriss@gmail.com}

\usepackage{arxiv}

\usepackage[english]{babel}

\usepackage{cleveref}
\crefrangelabelformat{equation}{(#3#1#4--#5#2#6)}
\crefname{equation}{Eq.}{Eqs.}
\Crefname{equation}{Equation}{Equations}

\usepackage{amsthm}

\theoremstyle{definition}
\newtheorem{defn}{Definition}[section]

\newtheorem{prop}{Proposition}[section]
\newtheorem{case}{Case}

\usepackage[utf8]{inputenc} 
\usepackage[T1]{fontenc}    
\usepackage{hyperref}       
\usepackage{url}            
\usepackage{booktabs}       
\usepackage{amsfonts}       
\usepackage{nicefrac}       
\usepackage{microtype}      
\usepackage{lipsum}
\usepackage{graphicx}
\graphicspath{ {./images/} }

\usepackage{csquotes}

\newcommand{\sigkapplam}{{\kappa,\!\lambda,\!\sigma\!}}
\newcommand{\kaplam}{{\kappa,\!\lambda\!}}
\newcommand{\sigkap}{{\kappa,\sigma\!}}
\newcommand{\siglam}{{\lambda,\sigma}}

\newcommand{\blam}{b_\lambda}
\newcommand{\dotblam}{\dot{b}_\lambda}

\begin{document}
\maketitle

\begin{abstract}
    This paper develops a mathematical framework for building a position in a stock over a fixed period of time while in competition with one or more other traders doing the same thing. We develop a game-theoretic framework that takes place in the space of trading strategies where action sets are trading strategies and traders try to devise best-response strategies to their adversaries. In this setup trading is guided by a desire to minimize the total cost of trading arising from a mixture of temporary and permanent market impact caused by the aggregate level of trading including the trader and the competition. We describe a notion of equilibrium strategies, show that they exist and provide closed-form solutions.
\end{abstract}

\keywords{Trading, position-building, game theory, Euler-Lagrange, trading strategies, equilibrium, strategy selection under uncertainty}

\section{Introduction}

    Participants in financial markets, in particular stock traders, often find themselves wanting to own a target quantity of stock on or before a future date. A typical scenario is when there is a catalyst that is expected to cause a change in the price and a trader who would like to own a target amount of the stock prior to that date. In such a case the trader is often aware that there are other traders looking to purchase the stock over the same period of time for the same reason. The problem for each trader is that the collective pressure all traders put on the price of the stock and cost of trading can have a dramatic impact on the success of the strategy. If a given trader delays trading then other traders may "drive up" the price of the stock increasing the cost of acquisition.
    
    In simple terms, this paper develops a framework for reasoning about this situation through an analysis of the {\em best-response} a trader may take against the one or more other players' trading strategies; this best-response is arrived at through the analysis of the market impact all trading activity has on the stock. To address this challenge we develop a version of game theory where each trader's available set of actions are trading strategies themselves.
    
    This problem shares some features in common with {\em optimal liquidation} strategies, as in the Almgren-Chriss model \cite{almgren2001optimal} but has significant difference because total trading costs are driven by aggregate trading. 

    \subsection{Computational Methods and Plot Generation}

    The differential equations presented in this paper were solved using Wolfram Mathematica \cite{Mathematica}, while the plots were generated using a combination of Wolfram Mathematica and Python. These tools provided the necessary computational accuracy and flexibility to visualize the results effectively.
    \section{Overview}
    
    This paper is organized into several sections, each addressing key aspects of optimizing position-building strategies in competitive trading environments.

    \textbf{Section \ref{sec:preliminaries}} defines the key concepts used throughout the paper, including trading strategies, trajectories, trajectory morphology (the shape of trajectories) including a breakdown of the most common shapes.
    
    {\bf Section \ref{sec:trading-in-competition}} introduces the idea of position-building in competition using the concepts from the prior section.  It provides details concerning the market impact models used in the rest of the paper, including both {\em permanent} and {\em temporary} and gives a conceptual overview of how market impact influences position-building in competition. Finally, this section discusses the {\em total cost of trading} in competition and the importance of the balance between temporary and permanent impact. This leads to the important notion of $\kappa$-regimes.
    
    In \textbf{Section \ref{sec:preliminary-examples-trading}} we provide several motivating examples and calculations to provide a sense for the issues involve with position-building in competition. While this section is not strictly necessary in the sequel, the author believes it provides some insight into the motivations behind this paper.
    
    \textbf{Section \ref{sec:passive-position-building}} examines {\em passive position building strategies}, those that trade {\em as if} there is no competition. We use these to introduce our game-theoretic approach to understanding how traders should optimally select their strategies in response to competitors. This section gives an overview of the Almgren-Chriss algorithm, \cite{almgren2001optimal}, and uses it to motivate the types of passive strategies one might run across. The section reviews the Euler-Lagrange equation which is used extensively throughout. 
    
    {\bf Section \ref{sec:non-equi-best-best-response}} introduces {\em non-equilibrium best-response strategies}, the first use of game-theoretic notions in this paper. These are strategies that optimally minimize the total cost of trading when building a position in competition with an adversary trading a known strategy. This section derives the non-equilibrium best-response strategies to the three most important types of passive strategies, risk-averse, risk-neutral and eager. 
    
    {\bf Section \ref{sec:two-trader-equilibrium-strategies}} defines {\em equilibrium strategies} when two traders are trading in competition and derives a simple system of differential equations that determines equilibrium strategies. In simple terms, equilibrium strategies are those strategies that are the best-response strategies to one another, meaning that is $a$ and $b$ are in equilibrium then $a$ is 
    the best response to $b$ and $b$ is the best response to $a$. This section derives closed-form solutions for the two-trader equilibrium and explores the results with several examples and also explores how $\kappa$, the balance between permanent and temporary market impact, affects the shape of equilibrium strategies.
    
    {\bf Section \ref{sec:eqilibrium-many-small-traders}} derives another equilibrium, one in which rather than two adversaries of potentially different target trading sizes, there are many traders, each trading the same target size, all in competition with one another. We derive a {\em symmetric equilibrium}, one in which all traders trade the same strategy and are in pairwise-equilibrium, for this case and give closed-form solutions. Once again we explore how $\kappa$ impacts the shape of the resultant strategies.
    
    In {\bf Section \ref{sec:inverse-problem}} we describe the {\em inverse problem} which answers the question {\em if I trade a certain strategy, what is this strategy the best response to?} We give several examples and explore how this can be a useful diagnostic tool.
    
    In {\bf Section \ref{sec:strategy-selection-uncertainty}} we explore the role of {\em uncertainty} in the selection of optimal position-building strategies. We discuss how to select a position-building strategy when it is not known what your adversary's understanding of the market is. This section does not provide a complete theory, but does give an important motivating example through the complete analysis of an archetypal situation. The upshot is that strategy selection in these circumstances must be broadened from simply computing an equilibrium strategy, to computing a collection of equilibrium strategies, together representing the space of possible strategies that may arise when seen from your adversary's point of view. In this section we compute the example where there are two traders, one, called $A$, who is trading a single unit of stock and the other, called $B$, who is trading many units. The uncertainty arises because $B$ does not know whether $A$ believes there is one or many adversaries and $A$ will choose its strategy according to this belief. We show in this example that this uncertainty places $B$'s decision into a probabilistic framework and then the best-response strategy may be selected, for example, using mean-variance analysis.

    In \textbf{Section \ref{sec:strategy-selection}} suggests an approach to expand the work of Section \ref{sec:strategy-selection-uncertainty} and place it in a rigorous probabilistic framework by introducing the idea of {\em probabilistic strategy selection}. We provide a partially-worked example.

    In \textbf{Section \ref{sec:mis-estimation}}, we explore the impact of parameter mis-estimation, particularly focusing on the market impact parameter $\kappa$ in two-trader equilibrium strategies. Mis-estimating $\kappa$ can significantly affect the total cost of trading, as illustrated in Figure \ref{fig:mis-estimation-kappa}. We provide a numerical exposition of how variations in $\kappa$ influence costs across a range of values. Additionally, we present sensitivity analysis of the total cost of trading with respect to both $\kappa$ and $\lambda$, highlighting how errors in these parameters can lead to suboptimal strategies. The detailed results are captured in Tables \ref{tab:impact-results} and \ref{tab:impact-results-1}. Finally, we briefly discuss possible extensions involving holding risk, leaving more detailed analysis for future work.

    Finally, in \textbf{Section \ref{sec:two-trader-equi-with-risk-aversion}} we introduce risk-aversion into the study of position-building with competition. Specifically we propose equilibrium equations analogous to those in Section \ref{sec:two-trader-eq-strats} but add a risk-aversion term to the loss functions that penalizes the total volatility a trader's position holds during the course of building the position. We proceed to study various numerical examples to gain intuition regarding the relative important of the parameters involved.
    
\section{Preliminaries}
\label{sec:preliminaries}

    This section sets forth the precise definitions and terminology used throughout this paper. 

    \subsection{Trading strategies}

\begin{figure}[h!]
        \centering
        \includegraphics[width=0.4\linewidth]{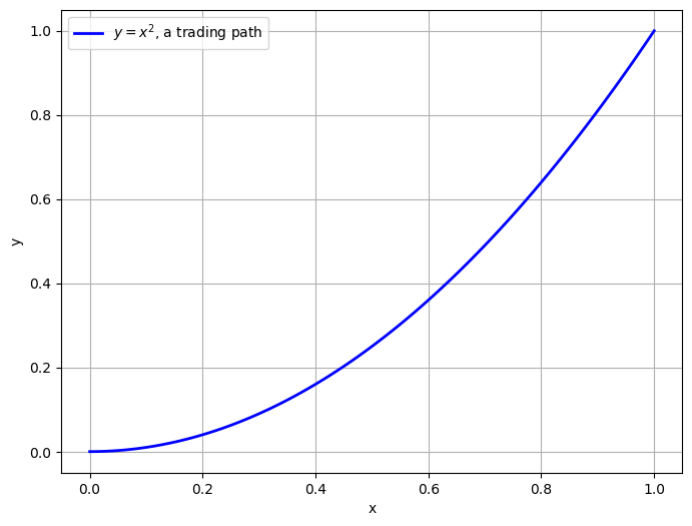}
        \caption{Example trading strategy with trajectory $y=x^2$}
        \label{fig:example-trading-path}
    \end{figure}

    A trading strategy is a mathematical description of the path that trading takes over a fixed interval of time, expressed in terms of the quantity of stock owned at each point in time. Put another way, it is the graph of a function showing the relationship between quantity in units of currency and time. In this paper, time is represented by $t$ and trading always starts at time $t=0$ and ends at time $t=1$. See Figure \ref{fig:example-trading-path}.
 
    \begin{defn}[Trading Strategy]
        A trading strategy (or, simply, {\em strategy}) is twice-differentiable function of time, say $x(t)$, that describes the units of stock held by a trader at each time $t$ between 0 and 1, inclusive. 
    \end{defn}  

    Trading strategies have various features that we will refer to throughout this paper and we enumerate them here:

    \begin{itemize}

        \item {\em Type:} the type of the trading strategy, references the purpose of the strategy and is defined formally in Definition \ref{def:strategy-type} below;

        \item {\em Start time and end time:} in the context of this paper every trading strategy has a definite start and end time. When not stated otherwise the start and end times will be $t=0$ and $t=1$, respectively;
        
        \item {\em Trajectory:} the trajectory of a trading strategy is the path it takes from start time to end time, often thought of as the graph of the strategy function; and

        \item {\em Shape:} the geometry of the trading trajectory, see Section \ref{sec:shapes} for more details.
    \end{itemize}

    We now formally define strategy type.

    \begin{defn}[Strategy type]
    \label{def:strategy-type}
    There are several important types of trading strategies and here is non-exhaustive list for a strategy $x(t)$:  
    \begin{itemize}
        \item {\em Liquidation}: Strategies for which $x(0)>0$ and $x(1)=0$, in other words a strategy that starts with a positive quantity of stock and ends with none;
        
        \item {\em Position-building}: Strategies for which $x(0)=0$ and $x(1)>0$;
        
        \item {\em Unit}: This is a technical for this paper in which a trader seeks to acquire a single unit of stock, usually in reference to an adversary who wishes to acquire a larger quantity; and
        
        \item {\em $\lambda$-Scaled}: Position-building strategies whose rate of trading is scaled by a constant $\lambda \ge 1$. By convention these are unit strategies that are scaled by a constantly factor $\lambda>0$. 
    \end{itemize}
    \end{defn}

    Note that scaled trading strategies represent strategies that have the "shape" of unit position strategies but which are scaled at each time $t$ by a fixed constant $\lambda>1$. This means that for a $\lambda$-scaled strategy $b(t)$, its trajectory is $\lambda \cdot b(t)$, while its {\em shape} is given by $b(t)$. We discuss strategy shape in the next section.

    \subsection{Trajectory morphology}
    \label{sec:shapes}

    It is sometimes useful to categorize trading trajectories morphologicallly, that is, according to their shapes.
    There are three basic shapes all of which have constant sign of second derivative:
    
    \begin{itemize}
        \item {\em Risk-neutral}: Strategies for which $\ddot x(t)= 0$ for all $t$. Risk-neutral position-building strategies are always of the form $\lambda\cdot t$ for some $\lambda>0$;
        
        \item {\em Risk-averse}: Strategies for which $\ddot x(t) > 0$ for all $t$; and 
        
        \item {\em Eager}: Strategies for which $\ddot x(t) < 0$ for all $t$.
    \end{itemize}

\begin{figure}[h!]
        \centering
        \includegraphics[width=0.333\linewidth]{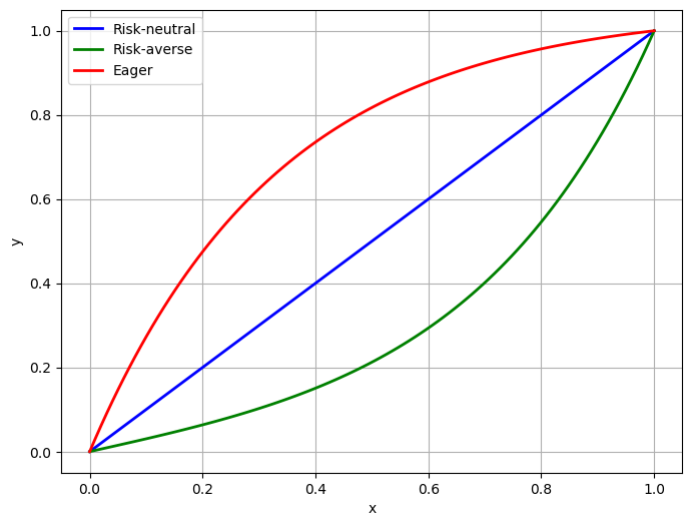}
        \caption{Visual depiction of risk-neutral, risk-averse and eager trading strategies}
        \label{fig:three-example-strategies}
    \end{figure}

    Figure \ref{fig:three-example-strategies} shows a visual description of the three types of position building strategy shapes. In subsequent sections we will discuss how these shapes relate to competitive position building.
    
    In addition to the above depicted shapes there are two other shape types that arise in practice.

    \begin{itemize}
        \item {\em Bucket:} bucket strategies acquire more than their target quantity immediately after the start time and then sell down to their target quantity as close to the completion time as possible; and

        \item {\em Barbell:} a barbell strategy buys a portion of its target quantity at the very start of trading and the remaining amount very close to the end of trading.
    \end{itemize}

\begin{figure}[h!]
        \centering
        \includegraphics[width=0.5\linewidth]{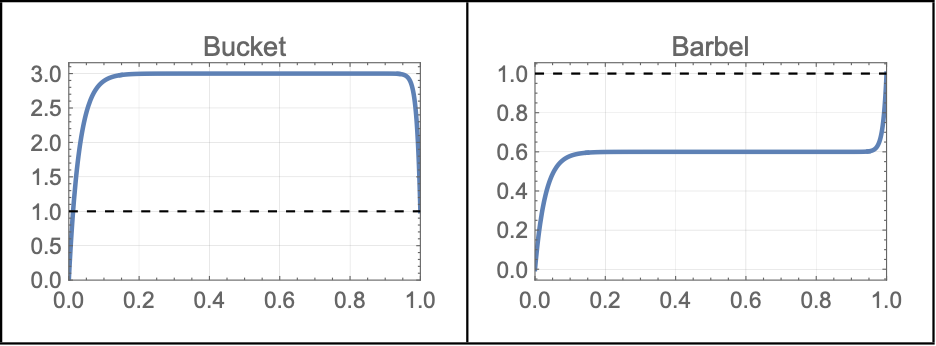}
        \caption{Example of bucket and barbell strategies. The bucket strategy shape acquires more than its target quantity almost immediately and then sells down to the target quantity as close to the completion time as possible. The barbell strategy by contrast acquires an intermediate fraction of the target quantity right away, holds steady and then acquires the rest as close to the completion time as possible.}
        \label{fig:bucket-and-barbell}         
    \end{figure}

    See Figure \ref{fig:bucket-and-barbell} for a visualization of these types.
    
    \subsection{Conventions and notation }
    \label{sec:sizing-conventions}

    For the remainder of the paper we will adhere to the following conventions and  notation.

    \begin{itemize}
        \item {\em Traders:} traders will be denoted by capital letters, e.g., $A$ and $B$. When not otherwise stated $A$ will denote a unit trader and $B$ a $\lambda$-scaled;

        \item {\em Strategies:} given a trader $A$, its associated strategy will refer to a function of time representing the strategy's trajectory. The associated strategies to a trader will be identified by the associated lower-case letter. For example the strategy trader by trader $A$ will be $a(t)$;

        \item {\em Twice-differentiable:} all strategies, say $a(t)$, will be twice-differentiable;

        \item {\em Unit position-building:} unless otherwise specifically stated, all strategy functions, say $a(t)$, will be unit position building strategies, that is, $a(0)=0, a(1)=1$; and

        \item {\em $\lambda$-scaling:} when a trader, say $B$, is trading a strategy that intends to acquire an arbitrary quantity of stock, $\lambda$, then we scale the strategy by $\lambda$, and thus the trajectory of $B$'s $\lambda$-scaled strategy is $\lambda\, b(t)$.
    \end{itemize}

\section{Trading in competition}
\label{sec:trading-in-competition}

    In this section we lay the groundwork for studying {\em trading in competition}, the situation in which two or more traders are trading in the same stock over the same stretch of time. When this occurs each trader must contend with the impact of the others' trading. We call this situation {\em trading in competition}. If they are both building a position in the stock than the situation is generally referred to as {\em competitive position building}. 

    In general this paper considers how to optimally adapt to the situation where two or more traders are trading in competition. In order to study this we need a firm understanding of what makes the situation more complex than trading {\em solo}. The starting point is market impact. 

    \subsection{Market impact}
    \label{sec:market-impact}

    Market impact refers to the impact on the price of a stock that is a direct consequence of trading activity. In the context of execution of trading strategies one can think of there being a vast pool of traders whose net impact of the price of a stock is essentially zero and a small pool, {\em one or more}, of traders directly trading so that {\em on net} the price of the stock moves in the direction of trading by this small pool. If the traders are buying, the price moves up, if they are selling, it moves down. We collect these ideas into a single definition here.

    \begin{defn}[Concerted and noise trading]
    \label{def:concerted-trading}
    At any given moment we divide the trading activity in a stock into two types:

    \begin{itemize}
        \item {\em Noise trading:} trading that is happening persistently as background activity by many traders and which is expected to have no net impact on price; and

        \item {\em Concerted trading:} trading that is being conducted by a few traders in the same direction (i.e., buying or selling), usually over a relatively brief period of time and whose net impact is expected to move the price in the direction of trading (e.g., up if the direction of trading is buying).
    \end{itemize}
    \end{defn}

    With this out of the way, we move on to discuss market impact. As market impact has an out-sized effect on both market function and investment performance, it is a subject of enormous practical and academic interest. In the context of this paper, we are concerned with both how immediate demand for a stock changes the price at which the stock may be purchased, and also how persistent demand for the stock over time causes an on-going change in the price. These forms of price change are respectively referred to as {\em temporary} (also referred to as {\em transient}) and {\em permanent} market impact. 

    The study of market impact in the academic literature is  vast and heterogeneous. There are strictly formal models, both linear and non-linear, models of limit and market order dynamics and models concerning market impact estimation. See \cite{almgren2001optimal}, \cite{almgren2001optimal}, \cite{gatheral2011optimal}, \cite{gatheral2013dynamical}, \cite{gatheral2010no}, \cite{obizhaeva2013optimal}, \cite{hautsch2012market}, \cite{zarinelli2015beyond} and \cite{almgren2005direct}\footnote{There are many models of market impact and methods for estimation of it, and the list provided is surely a small subset. See for for example the excellent overviews in  \cite{gatheral2010three}, \cite{webster2023handbook} and \cite{donnelly2022optimal}.}. In this paper we will use linear models for both temporary and permanent market impact following \cite{almgren2001optimal} and define them next.

    \begin{defn}[Temporary market impact]
    \label{def:temp-mkt-impact}
    The cost of trading relative to the prevailing market price of a stock as trading commences. Temporary impact affects the price of a stock only at the moment of trading and has not bearing on the future price. It is measured as the difference between the prevailing price at the time of an order and the average execution price of the order. The fact that there is temporary impact at all is presumed to arise from a premium being charged for {\em immediacy} of order execution.
    \end{defn}

    In some sense the polar opposite of temporary market impact is permanent market impact, defined as follows:

    \vskip 10pt

    \begin{defn}[Permanent market impact]
    \label{def:perm-mkt-impact}
    Is the change in the price of a stock due to persistent trading over time. The impact in this case is measured relative to the price of the stock when trading commences, and persists so long as concerted trading is taking place. 
    \end{defn}

    \subsection{Market impact while trading in competition}

    While there has been a great deal of academic work concerning the optimal execution of trading strategies in the presence of market impact, there has been scant discussion of trading in competition. As such we take a moment here to discuss the specific mathematical formulation of temporary and permanent market impact with and without competition that will be used in the remainder of this paper.

    To begin, we note that in the context of optimal trading strategies, whether with or without competition, the objective relates to minimizing the cost of trading relative to the price that {\em would have been} in the absence of trading. In this context one has to imagine a counterfactual price path {\em unperturbed} by concerted trading (see Definition \ref{def:concerted-trading}). Then one has to view the {\em implementation cost} of the strategy as the difference between trading at actual market prices and trading at the unperturbed prices.

    With this said we set forth the linear models for temporary and permanent market impact with and without competition. Let $A$ and $B$ be traders, trading $a(t)$ and $b(t)$ in competition. Then we have the following mathematical formulations of temporary and permanent market impact defined conceptually in Definitions \ref{def:temp-mkt-impact} and \ref{def:perm-mkt-impact} respectively.

    \vskip 8pt

    {\bf Temporary impact for $A$ without competition:} the impact on the price paid when $A$ is trading without competition is proportional to $\dot a(t)$ at time $t$ and the total cost is $\dot a^2(t)$;

    {\bf Permanent impact for $A$ without competition:} the impact on the price paid at at time $t$ when $A$ traded without competition starting at time $t=0$ is proportional to $a(t)$ and the total price paid is $a(t)\cdot \dot a(t)$;
        
    {\bf Temporary impact for $A$ competing with $B$:} the impact on the price paid when $A$ is trading in competition with $B$ is proportional to $\dot a(t) + \dot b(t)$ at time $t$ and the total cost is $(\dot a(t) + \dot b(t)) \cdot \dot a(t)$;

    {\bf Permanent impact for $A$ competing with $B$:} the impact on the price paid at at time $t$ when $A$ is trading in competition with $B$ is proportional to $a(t) + b(t)$ at time $t$ and the total cost is $(a(t) + b(t)) \cdot \dot a(t)$;

    \vskip 8pt

    The costs enumerated above are all stated as only {\em up to a proportionality constant} for two reasons. First, when searching for optimal trading strategies, knowing a quantities up to a constant multiple is all that is necessary, as with any optimization problem. Second, perhaps more importantly, what matters is how big permanent impact is {\em relative to} temporary impact.

    \subsection{Total Cost of trading}
    \label{sec:total-cost-of-trading-in-comp}
    
    We now set forth equations that determine the total cost of trading when two traders, $A$ and $B$, are trading in competition. These values will be computed {\em only up to a constant}. To do this we set the {\em temporary impact} constant of proportionality above to one and the {\em permanent impact} constant of proportionality to a value $\kappa>0$. We call $\kappa$ the {\em market impact coefficient} and we will use it throughout this paper. We now write:
    
    \begin{align}
        \label{eq:inst-cost-of-trading}
        \text{Instantaneous total cost of trading for $A$} & \propto  (\dot{a} + \dot{b}) \,\dot{a} + \kappa (a + b)\, \dot{a} \\
        \text{Instantaneous total cost of trading for $B$} & \propto  (\dot{a} + \dot{b}) \,\dot{b} + \kappa (a + b)\,  \dot{b} 
    \end{align}

    Using \cref{eq:inst-cost-of-trading} it is easy to see that the average cost of trading from time $t=0$ to $t=1$ must be given by the integral of those expressions and that the aim of a trader acquiring a position is to minimize the total cost of trading. In particular the job of trader $A$ wishes to minimize the total cost of trading in competition given as follows. Let $\mathcal{S}$ be the set of position-building strategies that are twice-differentiable:

    $$
    \mathcal{S} = \{ a:[0, 1]\to \mathbf{R} \,|\, a(0) = 0, a(1)=1, \ddot a\, \text{exists} \}
    $$

    and let $b(t)$ be the strategy trading competitively with $A$. We wish to place into a mathematical $A$'s goal in trading which is to minimize the cumulative total cost of trading between the start and end of building the position.
    
    \begin{equation}
        \label{eq:trading-program-a}
        \underset{a \in \mathcal{S}}{\text{minimize}} \, \int_0^1  \big(\dot a(t) + \dot b(t) \big) \dot a(t) + \kappa \big(a(t) + b(t)\big) \dot a(t)\, \text{dt}
    \end{equation}

    namely this trader would like to minimize the total cost of acquiring the position. Analogously, trader $B$ solves the problem:

    \begin{equation}
        \label{eq:trading-program-b}
        \underset{b \in \mathcal{S}}{\text{minimize}} \, \int_0^1  \big(\dot a(t) + \dot b(t) \big) \dot b(t) + \kappa \big(a(t) + b(t)\big) \dot b(t)\, \text{dt}
    \end{equation}

    In addition to solving \crefrange{eq:trading-program-a}{eq:trading-program-b} individually, we will also be interested in solving them jointly when the quantities they trade are different from one another. 

    \subsection{The balance of temporary and permanent impact: market imppact regimes}
    \label{sec:kappa-regimes}

    Much of the work in this paper is devoted to analyzing the properties of the cost functions described in Section \ref{sec:total-cost-of-trading-in-comp} and its influence on the shape of trading strategies in competition. To fully analyze this we start by refining our understanding of $\kappa$ in the total cost of trading formulas \cref{eq:inst-cost-of-trading}.

    The constant $\kappa$ in \cref{eq:inst-cost-of-trading} defines the relative quantity of transaction cost that arises from permanent versus temporary market impact. This quantity turns out to be the single most important quantity in analyzing trading in competition. We note this here because in optimal execution problems, see e.g. Section \ref{sec:almgren-chriss-review} and in particular \ref{sec:perm-impact-no-effect}, the presence of permanent impact has no effect on optimal execution strategies and is therefore ignored.

    By contrast, the presence of permanent market impact in competitive position-building problems is crucially important and the best way to view it is by means of {\em $\kappa$ regimes} which we explain now.

    \vskip 10pt 

    \begin{defn}[Kappa regimes]
    Kappa regimes refers to how the value of $\kappa$, the determination of the relative contribution of temporary and permanent market impact, influences the cumulative cost functions defined in \crefrange{eq:trading-program-a}{eq:trading-program-b}. We identify two broadly different $\kappa$-regimes according to whether total loss is dominated by temporary or permanent market impact:

    \begin{itemize}
        \item {\em Temporary impact dominated, $\kappa<1$}: in this regime, particularly when $0<\kappa \ll 1$, trading costs are dominated by temporary market impact costs and traders do not concern themselves as much with a persistent rise in the price of the stock and concern themselves more with minimizing temporary cost. As a consequence they tend to trade strategies that are closer to {\em risk-neutral} in shape, as in Section \ref{sec:shapes}; and

       \item {\em Permanent impact dominated, $\kappa>1$}: in this regime, particularly when $\kappa \gg 1$, trading costs are dominated by the permanent market impact costs and traders are concerned with "trading ahead" of their competition in order to buy before the price rises. As a consequence the shape of their strategies tend to be eager, as described Section \ref{sec:shapes}.
    \end{itemize}        
    \end{defn}

\section{Preliminary examples and motivation for optimal strategies in competition}
\label{sec:preliminary-examples-trading}
\label{sec:preliminary-examples}
    In order to building intuition we digress into two examples that will help to motivate what follows.

    
    \subsection{Example calculation}

    In order to clarify the the exact nature of the trading costs, an example will be useful. Suppose $B$ is trading the $\lambda$-scaled risk-neutral strategy $b(t)$: 

    $$   b_\lambda(t) = \lambda \cdot t$$

    and $A$ is trades a unit parabolic strategy {\em in competition} with $b(t)$:

    $$ a(t) = \frac{t(t-c)}{1 - c} $$

    In this case the rate of trading of $a(t)$ at time $t$ is given by:

    $$ \dot a(t) = \frac{1}{1 - c} \left( 2t - c \right) $$

    and the instantaneous cost of trading for $a$ at time $t$ is \( \text{cost}(t) \):

    $$
   \text{Cost}(t) = \big( \dot a(t) + \dotblam(t) \big) \cdot \dot a(t) + \kappa \cdot \big( a(t) + \blam(t) \big) \cdot \dot a(t)
    $$

    where $\dot b(t)=\lambda$. Then the cumulative cost function from time $0$ to $t$ is given by:

   $$
   \text{Cumulative cost}(t) = \int_0^t \big( \dot a(s) + \dotblam(s) \big) \cdot \dot a(s) + \kappa \cdot \big( a(s) + \blam(s) \big) \cdot \dot a(s) \, \text{ds} $$

\begin{figure}[ht]
        \centering
        \includegraphics[width=0.666\linewidth]{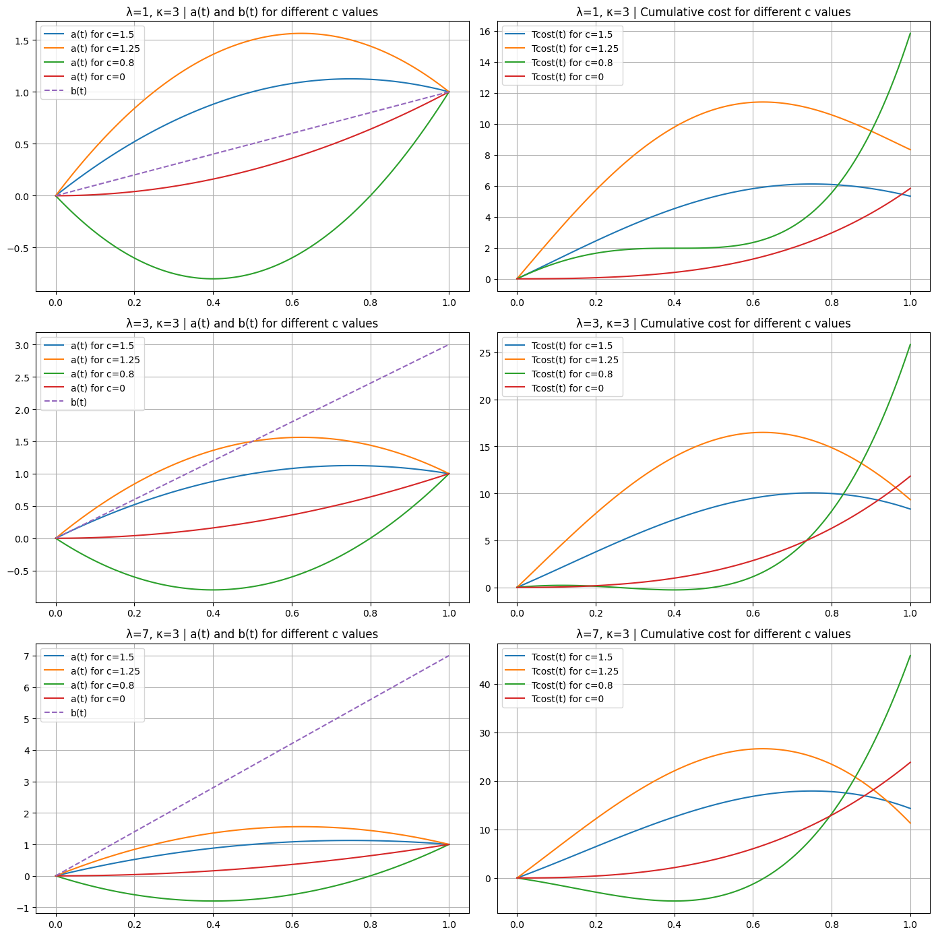}
        \caption{Example total costs for $a(t)$ in competition with $\blam(t)=t$ for $a(t) = \frac{1}{1-c}(2t - c)$. The left-hand column shows parabolic strategies $a(t)$ are denoted by solid lines and are all unit strategies with a variety of shapes from very risk-averse to very eager.  The dotted lines in the right-hand column represent the risk-neutral strategy $\blam(t)$ and as we move down the rows its $\lambda$-scaling increases. Each plot in the right-hand column shows total costs of the strategies. One can see from the plots that the market impact caused by the $B$ makes risk-averse strategies the most costly as $B$ pushes the price up as time goes on. On the other hand when $A$ trades eagerly they can profit from over-buying the stock at lower prices and selling for a profit later. This is reflected, for example, in the most eager strategy's cumulative cost peaking approximately sixty percent of the way through and then decreasing.}
        \label{fig:cumulative-cost-analysis}
    \end{figure}

   Figure \ref{fig:cumulative-cost-analysis} demonstrates the cumulative costs of a variety of strategies.

    \subsection{Motivating understanding the mix of temporary and permanent impact}

\begin{figure}[H]
        \centering
        \includegraphics[width=0.5\linewidth]{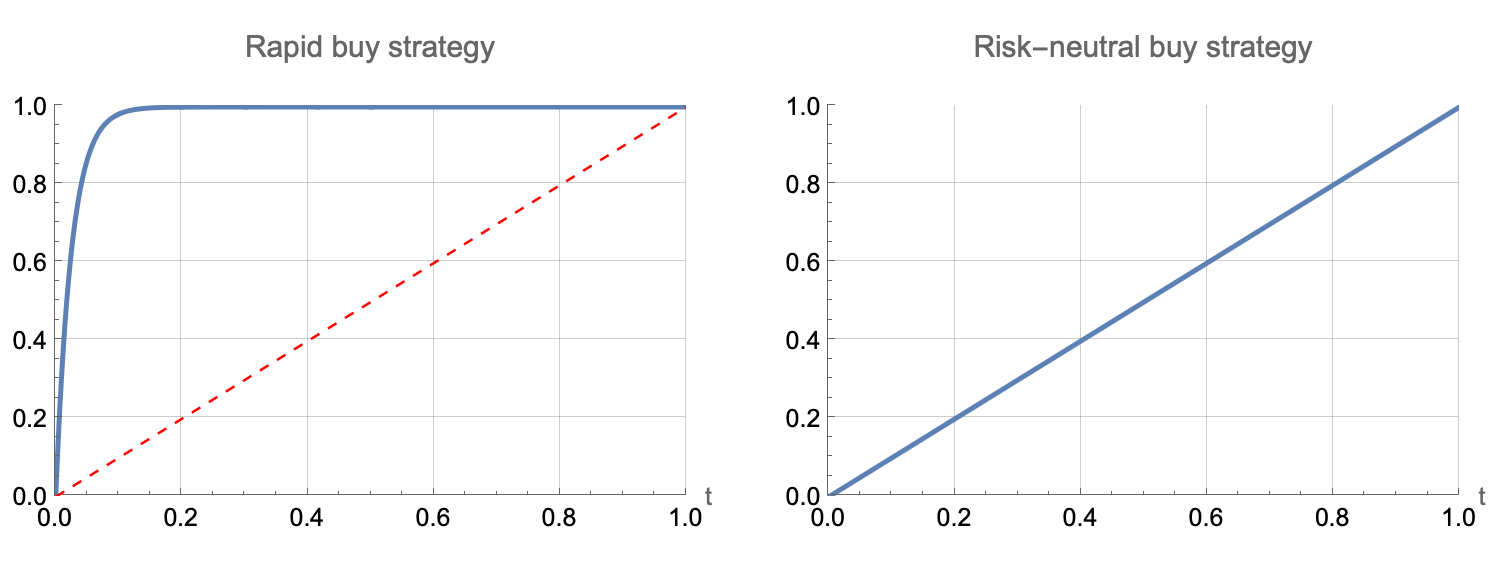}
        \caption{Rapid-buy and risk-neutral strategies with $B$'s risk-neutral strategy in dashed-red in the left plot.}
        \label{fig:rapid-buy-and-risk-neutral}
    \end{figure}

    In order to get a feel for the problem domain of trading in competition we start with the following problem. Suppose $A$ is trading in competition with $B$ and $A$ knows that $B$ is going to trade the strategy $b(t)=t$, a simple risk-neutral strategy. Further, $A$ only has two strategy options available, a rapid-buy and risk-neutral strategy respectively, as shown in Figure \ref{fig:rapid-buy-and-risk-neutral}.

    The first question we ask is, how should trader $A$ go about choosing between various strategies.
    
    To get at an answer we start by observing that the primary concern for a trader regarding permanent impact is that as $B$ purchases the stock the price will remain higher throughout the buy-program and subsequent prices. In the absence of permanent impact, the trader's only concern is minimizing temporary impact and the best way to do this in the case where $B$ is trading the risk-neutral strategy is to spread trades out as much as possible which clearly means trading the risk-neutral strategy as well (Figure \ref{fig:rapid-buy-and-risk-neutral} right plot).

    At the other extreme when there is no temporary market impact, the only concern is to avoid paying higher prices for the stock as $B$ pushes the price up during the course of buying. In this case since there is no penalty for a large quantity of purchases over a short period of time, the correct approach is to purchase all of the target quantity (one unit) immediately, and hence implement a rapid buy strategy (Figure \ref{fig:rapid-buy-and-risk-neutral} left plot).

\begin{figure}[h]
        \centering
        \includegraphics[width=0.85\linewidth]{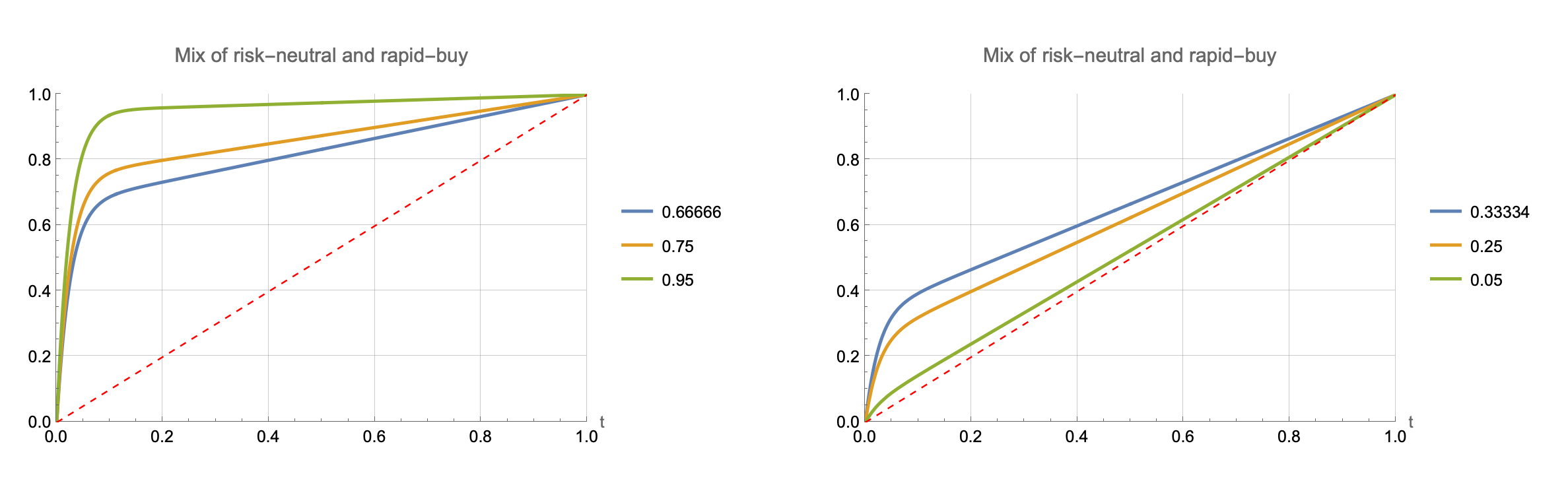}
        \caption{Convex combination of the rapid-buy and risk-neutral strategies.}
        \label{fig:mix-risk-neutral-and-rapid-buy}
    \end{figure}

    We see that in the two extreme cases of no temporary or no permanent impact it is fairly straightforward how to proceed when $B$ is trading a risk-neutral strategy. But what if both temporary and permanent impact are present, as in Section \ref{sec:kappa-regimes}? In this case we would need to implement more complex strategies. One possibility might be to estimate the $\kappa$ representing the mix of permanent and temporary impact as in Sections \ref{sec:total-cost-of-trading-in-comp} and \ref{sec:kappa-regimes} and then form a linear combination of a rapid-buy and a risk-neutral program. This is depicted in Figure \ref{fig:mix-risk-neutral-and-rapid-buy}. Not only is this not an optimal approach, it also does not indicate what to do if $B$ is trading a $\lambda$-scaled strategy and not targeting one unit of stock. It also misses the mark for what to do if $B$ is not trading a risk-neutral strategy.

    To provide complete answers to these questions we need a better framework that builds strategies that are optimal responses to other strategies taking into account how and how much they are trading. We will develop these soon enough but next we do a small example calculation to get a feel for how costs are calculated in practice.

    \section{Passive position-building strategies}
    \label{sec:passive-position-building}

    In this section we study {\em passive} position-building strategies, those strategies for which a trader builds a position without taking into account the actions of other traders (see section \ref{sec:preliminary-examples-trading}. We use the Almgren-Chriss algorithm to derive optimal {\em passive} position-building strategies for risk-averse traders in the absence of competition. This is not strictly necessary in the sequel but provides context for how the analysis of non-competitive trading differs from competitive trading. 

    \subsection{Almgren-Chriss optimal execution review}
    \label{sec:almgren-chriss-review}

    In this section we review the Almgren-Chriss algorithm \cite{almgren1997optimal} and \cite{almgren2001optimal}\footnote{The algorithm was originally developed while Chriss was working in the Institutional Equity Division at Morgan Stanley, while working on the program trading desk from 1996-1997 and published in expository form in \cite{almgren1997optimal}.}. Almgren-Chriss was developed to tackle the problem of buying or selling a large quantity of a stock or portfolio that is too large to be executed in one trade. The problem in that paper was to balance temporary market impact costs that increase with the rate of trading and the risk inherent in holding the stock. The paper also considers permanent impact costs but in this framework it is an artifact of the model that permanent impact costs have no bearing on the optimal trading strategy. We will review this next.

    The setup In Almgren-Chriss is that at time $t=0$ a trader owns a certain quantity of and seeks to {\em liquidate} it entirely by time $t=T$. The liquidation strategy's trajectory is described by a twice-differentiable function $x(t)$; the trader wishes to minimize some combination of temporary impact cost and risk as described by the following loss function\footnote{The function could easily be called a cost function and be denoted $C$ but in keeping with the Euler-Lagrange equation we stick with $L$.}:

    \begin{equation}
    \label{eq:loss-function-alm-chr}
        L(t; x) = \dot x^2 + \lambda \sigma^2 x^2
    \end{equation}

    This is a function of time, $t$, and we note its dependence on the strategy $x$ as well. The strategy $x(t)$ that has minimal cost among all strategies is called the {\em optimal strategy}. To find the minimal strategy is a straightforward application of the Euler-Lagrange equation\footnote{The Euler-Lagrange equation is ubiquitous, but see \cite{gelfand2000calculus} for a good overview.}. Then the Euler-Lagrange equation is used to find the function \( x(t) \) that minimizes a functional defined by means of the loss function \( L(t, x(t), \dot{x}(t)) \), where \( \dot{x}(t) \) denotes the derivative of \( x(t) \) with respect to \( t \), the Euler-Lagrange equation is given by:
    
    \begin{equation}
        \label{eq:euler-lagrange}
        \frac{\partial L}{\partial x} - \frac{\mathrm{d}}{\mathrm{d}t} \left( \frac{\partial L}{\partial \dot{x}} \right) = 0
    \end{equation}
    
    Below is a brief explanation of the notation:
    
    \begin{itemize}
        \item \( \frac{\partial L}{\partial x} \) is the partial derivative \( L \) with respect to \( x(t) \);

        \item  \( \frac{\partial L}{\partial \dot{x}} \) is the partial derivative of \( L \) with respect to \( \dot{x}(t) \); and
        
        \item \( \frac{\mathrm{d}}{\mathrm{d}t} \left( \frac{\partial L}{\partial \dot{x}} \right) \) is the total derivative with respect to time of \( \frac{\partial L}{\partial \dot{x}} \). 

    \end{itemize}

    Recall a position-building strategy $x(t)$ is one for which $x(0)=0$ and $x(1)=1$. We start by deriving an optimal position-building strategy with {\em no competition}. In this setup we find an extremum strategy for the loss function $L$ in \cref{eq:loss-function-alm-chr} and obtain the following  differential equation:
    
    \begin{equation}
        \label{eq:almgren-chriss-diffeq}
        \ddot x - \lambda\sigma^2 x = 0
    \end{equation}
    
    with boundary condition $x(0)=0, x(1)=1$. We solve for $x(t)$ for the given boundary conditions $x(0)=0$ and $x(1)=1$\footnote{We wil generally place the boundary conditions for these sorts of differential equations directly in-line with the equation itself.} to obtain 
    
    $$ 
    x(t) = \frac{\sinh(\sigma \sqrt{\lambda} \, t)}{\sinh(\sigma \sqrt{\lambda})}
    $$
    
    To simplify notation we absorb $\sqrt{\lambda}$ into $\sigma$, thus making $\sigma$ represent a kind-of {\em risk-aversion}-scaled volatility and call the resultant strategy the {\em risk-averse position-building strategy}: 
    
    \begin{equation}
    \label{eq:risk-averse}
    x(t) = \frac{\sinh(\sigma  \, t)}{\sinh(\sigma )}
    \end{equation}
    
\begin{figure}[H]
        \centering
        \includegraphics[width=0.45\linewidth]{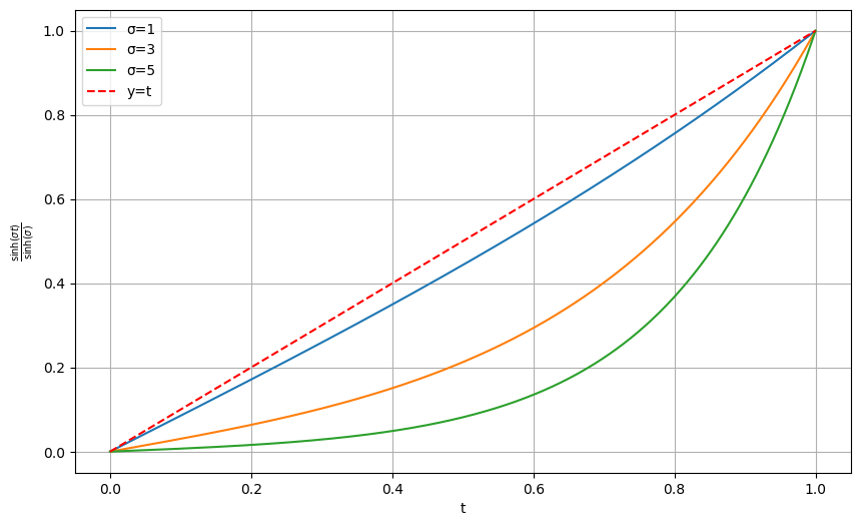}
        \caption{Risk-averse position-building strategies for various values of $\sigma$.}
        \label{fig:risk-averse-posn-building}    
    \end{figure}

    Figure \ref{fig:risk-averse-posn-building} shows an example risk-averse position-building strategy.

    \subsection{Permanent impact has no effect on optimal execution}
    \label{sec:perm-impact-no-effect}

    Suppose we augment the loss function \cref{eq:loss-function-alm-chr} with permanent impact so that
    
    \begin{equation}
        L(x) = \dot x^2 + \kappa \cdot x \cdot \dot x + \lambda \sigma^2 x^2
    \end{equation}
    
    then applying the Euler-Lagrange equation \cref{eq:euler-lagrange} we obtain

    \begin{align*}
        \frac{\partial L}{\partial x} &=  \kappa \dot x + 2\lambda\sigma^2 x \\
        \frac{\partial}{\partial t}\frac{\partial L}{\partial \dot x} &= 
            2 \ddot x + \kappa \cdot \dot x  
    \end{align*}

    and so from the Euler-Lagrange theorem \cref{eq:euler-lagrange} we obtain $\ddot x - \lambda\sigma^2 x = 0$. This implies that the differential equation describing $x$ is given by 

    \begin{equation}
        \ddot x - \lambda \sigma^2 x 
    \end{equation}

    precisely matching \cref{eq:almgren-chriss-diffeq}, thus confirming that the presence of permanent impact has no influence on the shape of an optimal execution strategy.

\section{Non-equilibrium best-response strategies}
\label{sec:non-equi-best-best-response}

    The previous section provides the building blocks of a game theory framework whose actions are drawn from the space of trading strategies and whose payoffs are described in terms of total cost of trading. In this setup there are two traders $A$ and $B$ in competition to buy the same stock starting at time $t=0$ and completing at time $t=1$. Both traders trade position-building strategies with trader $A$ trading a {\em unit strategy} and trader $B$ a $\lambda$-scaled strategy with $\lambda>1$. This setup corresponds to the notion of {\em differential games} and in particular {\em open-loop} games, as in \cite{isaacs1999differential}.
    
    To fashion this into a game, $A$ and $B$ simultaneously choose their strategies immediately prior to $t=0$ and agree to adhere perfectly to their respective strategies during the interval of time from $t=0$ to $t=1$. To begin our investigation we start by understanding how $B$ should trade knowing what strategy $A$ will trade. In the next few sections we derive {\em non-equilibrium best-response} strategies to particular strategies. The setup is as follows:

    \begin{itemize}
        \item Traders $A$ and $B$ are trading the same stock over the same period of time with $B$ trading a $\lambda$-scaled strategy;

        \item Trader $B$ is trading {\em passively} (see Section \ref{sec:preliminary-examples-trading}); and
        
        \item Trader $A$ knows precisely what $B$'s strategy is.
    \end{itemize}

    We derive $A$'s optimal strategy in light of $B$'s which in keeping with game-theoretic terminology we call the {\em best-response strategy} to $B$. To emphasize that the best-response strategy is trading versus a passive trader, we call the results {\em non-equilibrium} strategies and call trader $B$ an {\em adversary}.
    
\subsection{Best-response strategy to a risk-averse adversary}
\label{sec:best-response-risk-averse}

    Suppose trader $B$ follows a $\lambda$-scaled risk-averse position-building strategy as in \cref{eq:risk-averse}. If trader $A$ is completely aware of $B$'s strategy and wants to acquire one unit of the same stock over the same period of time, then what is the optimal strategy for trading $A$ to follow? To compute this we start by defining a loss function for $a(t)$ as follows:
    
    \begin{equation}
        \label{best-response-risk-averse}
        L(a, b) = \left(\dot a + \lambda \dot b\right) \cdot \dot a + \kappa \left( a + \lambda b\right) \cdot \dot a
    \end{equation}
    
    To explain this, start by noting the first component of the loss function is the product of $A$'s rate of trading and the total rate of trading of $A$ and $B$, where $B$'s trading is scaled by $\lambda$. The first component therefore represents the temporary market impact trader $A$ will experience trading in competition with $B$. Unlike in Almgren-Chriss the temporary impact includes the impact of both traders, reflecting the competitive nature of the trading.

    The second component of \cref{best-response-risk-averse} represents {\em permanent} market impact, where $\kappa$ is the fraction of total trading that is retained in the price of the stock. As with temporary impact, the permanent impact is determined by the trading of both $A$ and $B$. Applying the Euler-Lagrange equation yields

    \begin{align}
        2\ddot a + \lambda \ddot b + \kappa \dot a + \kappa\lambda \dot b  -\kappa \dot a &= 0 \quad \implies\nonumber\\
        \ddot a = - \frac{\lambda}{2} ( \ddot b + \kappa \dot b)\label{eq:br-basic}
    \end{align}
        
    We call \cref{eq:br-basic} the {\em best-response} equation for trader $A$ responding to a single adversary $B$ with known $\lambda$-scaled strategy $b(t)$. Now substituting \cref{eq:risk-averse} for $b$ we obtain 
    
    \begin{align*}
    a(t) &= - \frac{\lambda}{2} ( b + \kappa \int b) + wt + z\\
      &= - \frac{\lambda}{2}
      \left( \frac{\sinh(\sigma  \, t)}{\sinh(\sigma )} + \kappa\int \frac{\sinh(\sigma  \, t)}{\sinh(\sigma )}\right) + wt + z \\
      &= - \frac{\lambda}{2}\left(
      \frac{\sinh(\sigma  \, t)}{\sinh(\sigma )} + \frac{\kappa}{\sigma}\frac{\cosh(\sigma  \, t)}{\sinh(\sigma )}
      \right) + wt + z 
    \end{align*}
    
    where the linear portion $wt + z $ can be used to satisfy the boundary conditions $a(0)=0$ and $b(0)=0$.  To simplify this further write $\xi = \kappa/\sigma$ and 
    
    $$
    q(t) = \frac{\sinh(\sigma  \, t)}{\sinh(\sigma )} + \xi\frac{\cosh(\sigma  \, t)}{\sinh(\sigma )}
    $$
    and we have
    
    \begin{equation}
        a(t) = -\frac{\lambda}{2} q(t) + wt + z 
    \end{equation}
    
    Solving for the boundary conditions we have
    
    \begin{align*}
        a(0) &= -\frac{\lambda}{2} q(0) + z \\
        a(1) &= -\frac{\lambda}{2} q(1) + w + z
    \end{align*}
    
    which finally implies $z=\lambda/2 q(0)$ and 
    
    $$
    w = 1+\frac{\lambda}{2} \left(q(1) - q(0) \right)
    $$
    
    Putting this together we arrive at $A$'s best-response to $B$'s risk averse position-building strategy:
    
    \begin{equation}
    \label{best-response-strategy-ra}
        a(t) = -\dfrac{\lambda}{2} q(t) + \big(1+\dfrac{\lambda}{2} (q(1) - q(0) )\big)\, t + \dfrac{\lambda}{2}q(0)
    \end{equation}

    \vskip 10pt

    {\bf Summary:} We can summarise the best-response to a risk-averse trader more neatly as follows. The best response to the $\lambda$-scaled risk-averse strategy $b(t;\siglam) = \frac{\sinh(\sigma t)}{\sinh(t)})$ is given by a strategy $a(t; \sigkapplam)$, where we explicitly signal the dependence on the risk-aversion parameter $\sigma$, scaling $\lambda$ and market impact coefficient $\kappa$:

    \begin{equation}
    \label{best-response-strategy-ra-neat}
        a(t; \sigkapplam) = \dfrac{\lambda}{2}(q_\sigkap(0)- q_\sigkap(t)) + \big(1+\dfrac{\lambda}{2} (q_\sigkap(1) - q_\sigkap(0) )\big)\, t 
    \end{equation}

    where $q_{\sigma, \lambda}(t)$ is an auxiliary function depending on $\sigma$ and $\kappa$: 

    \begin{equation}
        \label{eq:qt-fn}
        q_{\sigkap}(t) = \frac{\sinh(\sigma \, t)}{\sinh(\sigma)} + \frac{\kappa}{\sigma}\cdot \frac{\cosh(\sigma  \, t)}{\sinh(\sigma )}
    \end{equation}
    
    \vskip 10pt
    
\begin{figure}[h]
        \centering
        \includegraphics[width=0.666\linewidth]{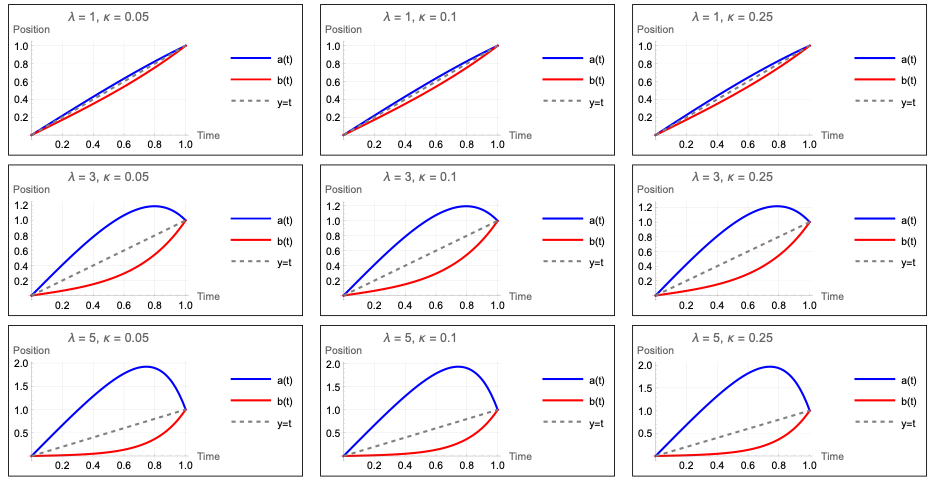}
        \caption{Best-response $a(t)$, in blue, trading in competition with a  risk-averse position-building strategy $b(t)$, in red, in a low $\kappa$ regime. The best-response strategies trade {\em eagerly} to get ahead of the pending price increases due to $B$'s trading.}
        \label{fig:br-risk-averse}
    \end{figure}

    {\bf Plots of the best-response to a risk-averse trader:} Figure \ref{fig:br-risk-averse} shows an example of the best-response to a risk-averse position-building strategy. In each plot the solid red line represents the strategy $b(t)$, the risk-averse trader's position-building strategy. In this situation the risk averse trader $B$ is simply trading without regard to the actions of trader $A$, while trader $A$ is trading optimally with respect to $B$. Notice that the blue lines are {\em trading ahead} of the risk-averse trader. The first row shows the situation where the risk-averse trader and the best-response target the same quantity (i.e., $\lambda=1$). The second column shows where $\lambda=5$, that is, the risk-averse trader is acquiring five times as much stock as the best-response. In this case the best-response is to over-buy and then sell to the risk-averse trader as they continue to buy. 

    In general we note that the best-response to a risk-averse strategy is {\em eager} in the sense of section \ref{sec:shapes}.

    \begin{figure}[H]
        \centering
        \includegraphics[width=0.75\linewidth]{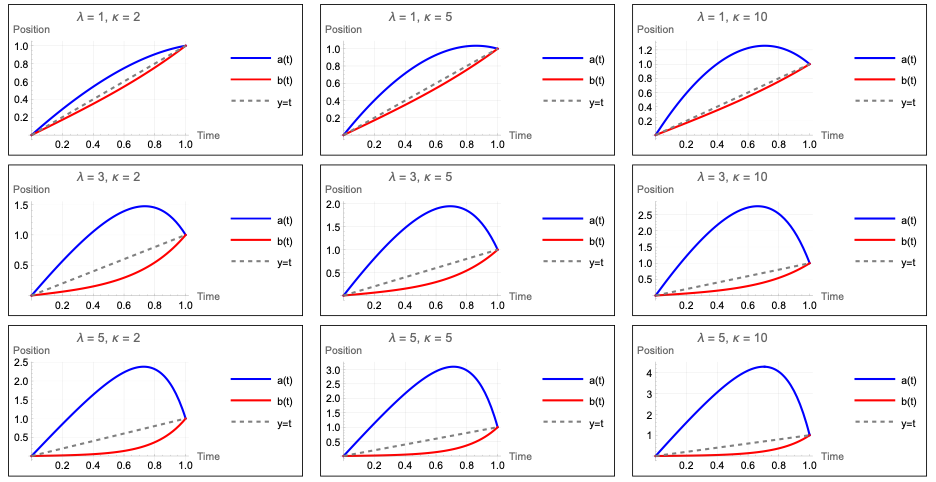}
        \caption{Best-response in competition with a risk-averse trader in a high $\kappa$ regime. These strategies trade versus the same risk-averse position builder in Figure \ref{fig:br-risk-averse} however now the $]\kappa$'s are all greater than one and the result is a significantly more eager strategy.}
        \label{fig:br-risk-averse-high-kappa}
    \end{figure}

\subsection{Best-response to a risk-neutral adversary}
\label{sec:best-response-risk-neutral}

    Suppose a competitor $B$ trades $\lambda$-scaled ($\lambda \ge 1$) risk-neutral strategy $b(t)=t$. The corresponding loss function is:
    
    \begin{equation}
        \label{eq:loss-fn-br-lambda-scaled}
        L = (\dot a + \lambda \dot b) \dot a + \kappa \cdot (a + \lambda b) \dot a
    \end{equation}
    
    and therefore from the Euler-Lagrange equation \cref{eq:euler-lagrange} the differential equation describing the unit best-response to a $\lambda$-scaled strategy is

    \begin{equation}
        \label{eq:diffeq-br-lambda-scaled}
        \ddot a = -\frac{\lambda}{2}( \ddot b + \kappa \dot b) = 0        
    \end{equation}
    
    from which we solve for $a(t)$ with boundary conditions $a(0)=0$ and $a(1)=1$.
    
    $$
    a(t) = \left(1 + \frac{\lambda \kappa}{4}\right) t - \frac{\lambda \kappa}{4} t^2 
    $$
    
    In this setup $A$ is reacting to a competitor $B$ who is trading a larger quantity.

    \begin{figure}[H]
        \centering
        \includegraphics[width=0.75\linewidth]{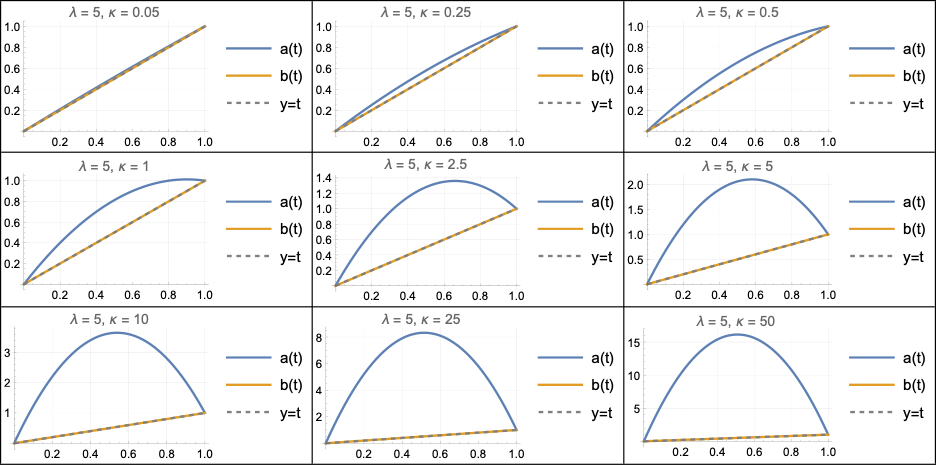}
        \caption{Best-response strategy to a risk-neutral trader for $\lambda=5$ in all cases and various $\kappa$ values.}
        \label{fig:best-response-to-linear}
    \end{figure}

    In general we note that the best-response to a risk-averse strategy is {\em eager} in the sense of section \ref{sec:shapes}.

\subsection{Best-response to an eager adversary}
\label{sec:best-response-eager}

    Finally we examine the case of the best-response strategy to an eager competitor. For this trader we use the $\lambda$-scaled strategy as follows:
    
    $$
    b(t; \siglam) = \frac{e^{-\sigma t} - 1}{e^{-\sigma} - 1}, \qquad\text{eager $\sigma$-scaled, $\sigma$ eagerness}
    $$
    
    as above we solve for the best-response strategy $a(t)$ using the differential equation \cref{eq:br-basic} with boundary conditions $a(0)=0, a(1)=1$ to obtain:

    \begin{equation}
        \label{eq:best-response-eager}
        b(t; \sigkapplam) = \frac{e^{\sigma  (-t)} \left(\lambda e^{\sigma } (\sigma -\kappa )-t e^{\sigma  t}
            ((\lambda +2) \sigma -\kappa  \lambda )+e^{\sigma+\sigma  t} (-\lambda  \sigma +\kappa  (\lambda
            -\lambda  t)+(\lambda +2) \sigma  t)\right)}{2\left(e^{\sigma }-1\right) \sigma }
    \end{equation}
    
\begin{figure}[h]
        \centering
        \includegraphics[width=0.3333\linewidth]{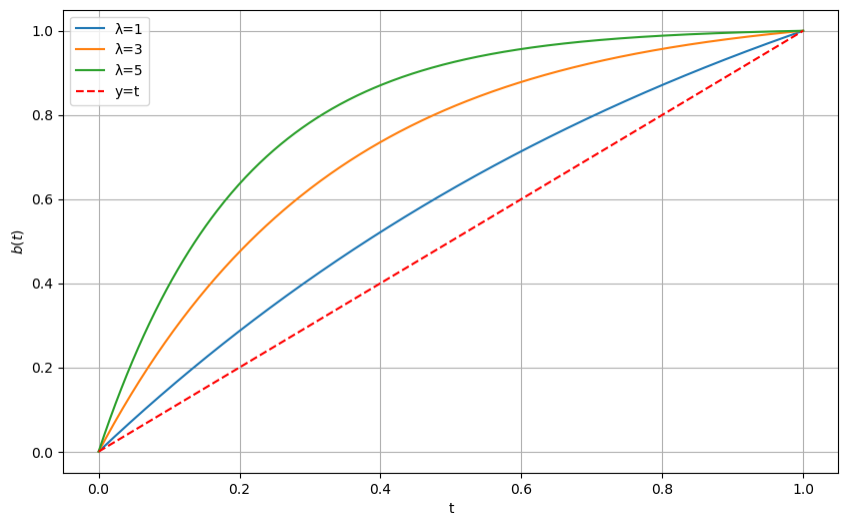}
        \caption{Eager position-building strategies.}
        \label{fig:eager-trader}
    \end{figure}

        
    Figure \ref{fig:eager-trader} shows an eager trading strategy for various values of $\lambda$.

\begin{figure}[H]
        \centering
        \includegraphics[width=0.75\linewidth]{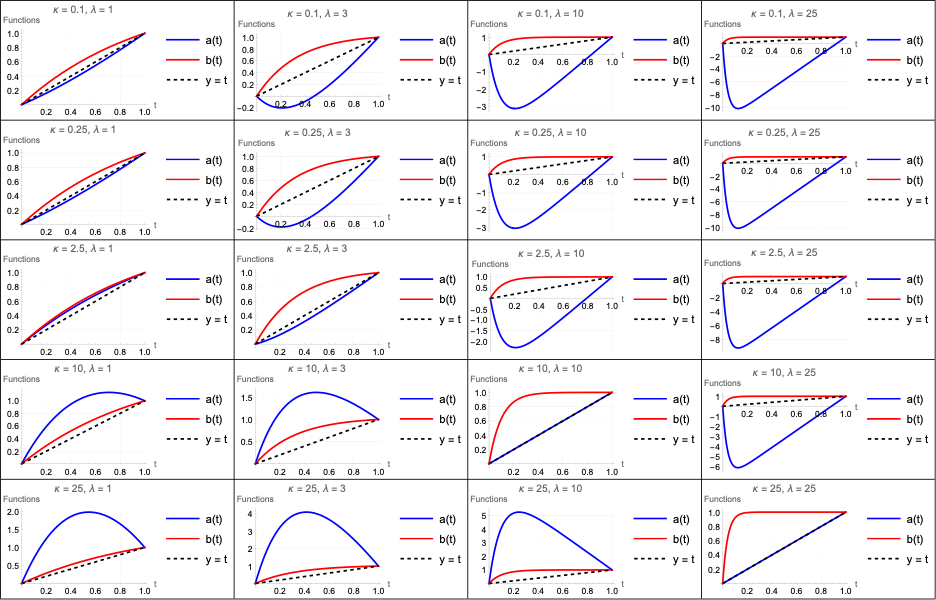}
        \caption{The best-response, in blue, to an eager position-building strategy, in red, in low- and high-$\kappa$ regimes. The eager strategy's shape, $b(t)$ remains fixed throughout but $\lambda$ increases from one to twenty-five moving from the left to the right column. Each row increases the size of $\kappa$ moving from low to high.}
        \label{fig:eager-compeitor-best-response} 
    \end{figure}

    Figure \ref{fig:eager-compeitor-best-response} shows the best-response to an eager trader for various values $\lambda$ (scaling) and $\kappa$ (permanent impact coefficient). We note that in general the best-response to an eager strategy is risk-averse. 

    Examining the plots we see something curious. The best-response to the eager trader counter-intuitively begins by selling short and then later re-buys the stock. Since the eager trader is rapidly pushing the price of the stock up, can this possibly be correct? The answer comes down to how the model {\em precisely} deals with market impact.

    \vskip 10pt

    \begin{table}[H]
    \centering
    \caption{The temporary and permanent impact costs of the best-response (first number) and risk-neutral (second number) strategies in competition with an eager strategy b(t) with eagerness $\sigma=4$, as depicted in Figure \ref{fig:eager-compeitor-best-response}.}
    \begin{tabular}{|c|c|c|c|} \hline
    \textbf{\( \lambda\), \(\kappa\)} & \textbf{Temp} & \textbf{Perm} & \textbf{Total} \\ \hline
    \(1, 0.10\) & \({2.12}, 2.0\) & \({0.12}, 0.11\) & \({2.24}, 2.11\) \\ \hline
    \(1, 0.25\) & \({2.10}, 2.0\) & \({0.30}, 0.27\) & \({2.41}, 2.27\) \\ \hline
    \(1, 2.50\) & \({1.99}, 2.0\) & \({2.84}, 2.70\) & \({4.82}, 4.70\) \\ \hline
    \(1, 10\) & \({2.81}, 2.0\) & \({8.72}, 10.82\) & \({11.53}, 12.82\) \\ \hline
    \(1, 25\) & \({10.14}, 2.0\) & \({8.64}, 27.05\) & \({18.78}, 29.05\) \\ \hline
    \(3, 0.10\) & \({2.63}, 4.0\) & \({0.39}, 0.27\) & \({3.02}, 4.27\) \\ \hline
    \(3, 0.25\) & \({2.60}, 4.0\) & \({0.96}, 0.66\) & \({3.56}, 4.66\) \\ \hline
    \(3, 2.50\) & \({2.93}, 4.0\) & \({7.82}, 6.64\) & \({10.75}, 10.64\) \\ \hline
    \(3, 10\) & \({15.08}, 4.0\) & \({7.76}, 26.57\) & \({22.84}, 30.57\) \\ \hline
    \(3, 25\) & \({90.40}, 4.0\) & \({-98.1}, 66.4\) & \({-7.8}, 70.4\) \\ \hline
    \(10, 0.10\) & \({-56.6}, 11\) & \({1.88}, 0.95\) & \({-54}, 12\) \\ \hline
    \(10, 0.25\) & \({-55.0}, 11\) & \({4.61}, 2.38\) & \({-50.33}, 13\) \\ \hline
    \(10, 2.50\) & \({-21.1}, 11\) & \({32.71}, 23.75\) & \({11.66}, 35\) \\ \hline
    \(10, 10\) & \({214.7}, 11\) & \({-48.29}, 95.0\) & \({166}, 106\) \\ \hline
    \(10, 25\) & \({1253}, 11\) & \({-1016}, 238\) & \({237}, 249\) \\ \hline
    \(25, 0.10\) & \({-580}, 26\) & \({5.51}, 2.5\) & \({-574}, 28\) \\ \hline
    \(25, 0.25\) & \({-563}, 26\) & \({13.5}, 6.1\) & \({-549}, 32\) \\ \hline
    \(25, 2.50\) & \({-245}, 26\) & \({91}, 61\) & \({-154}, 87\) \\ \hline
    \(25, 10\) & \({1582}, 26\) & \({-226}, 245\) & \({1356}, 271\) \\ \hline
    \(25, 25\) & \({8776}, 26\) & \({-3510}, 613\) & \({5267}, 639\) \\ \hline
    \end{tabular}
    \label{tab:temp-perm-strategy-costs}
    \end{table}
    
    In Table \ref{tab:temp-perm-strategy-costs} we show the cost of trading both the best-response and risk-neutral unit strategy trading in competition with a passive, eager trader for the values of $\kappa$ and $\lambda$ in Figure \ref{fig:eager-compeitor-best-response}. Beginning with the total column we see that the best-response strategy's total cost is indeed lower than the risk-neutral strategy's. For example, for $\lambda=1$ and $\kappa=0.1$ the values are very close but the best-response strategy's total cost was 2.09 versus the risk-neutral's of 2.11, while for $lambda=10$ and $\kappa=0.10$ the best-response strategy's total cost is significantly lower at -86.1 versus the risk-neutral strategy's total cost of 12.

    To understand what is going on we look at the temporary and permanent impact columns. These columns show the calculated contribution of temporary and permanent impact to the total cost of trading. Examining these it is straightforward to see that the best-response strategy benefits from lower (and in the case of large $\lambda$, {\em significantly} lower) temporary impact costs. Why is this the case? To arrive at the answer we have to carefully examine the handling of temporary impact in the market.

    \subsection{Temporary impact revisited}

    Recall from Section \ref{sec:market-impact} that temporary market impact relaxes instantaneously the moment a trade occurs and has no bearing on the future price of a stock. In non-competitive trading, the meaning of this is clear both when the trader is buying and when the trader is selling. In the case of a passive trader, the temporary impact cost imposed on a trader trading at an instantaneous rate of (say) $\dot x(t)$ at time $t$ is $\dot x^2(t) > 0$. In other words the cost is {\em always positive}.

    Compare this, however, to the situation in which a trader $A$ is trading $a(t)$ in competition with another trader, say $B$, trading $\lambda b(t)$. In this situation, the market impact cost to $A$ will depend on the {\em net} trading in the stock. It will be a profit (a negative cost) whenever $\text{sgn}(\dot a(t) + \lambda{\dot b(t)}) \ne \text{sgn}(\dot a(t))$. That is, whenever the trading direction of  $A$ and $B$ is {\em in aggregate} the opposite sign of $A$'s then $B$'s temporary impact will create a profit opportunity for $A$. We illustrate this in a small table:

    \begin{table}[h!]
    \centering
    \label{tab:impact_sign}
    \renewcommand{\arraystretch}{1.25}
    \begin{tabular}{|l|l|l|l|l|l|}
    \hline
    \textbf{A action} & \textbf{B action} & \textbf{A sign} & \textbf{A + B sign} & \textbf{A Impact Sign} & \textbf{A Impact} \\ \hline
    A buy             & B buy             & $\dot{a} > 0$   & $\dot{a} + \dot{b} > 0$ & $(\dot{a} + \dot{b}) \dot{a} > 0$ & Cost \\ \hline
    A sell            & B sell            & $\dot{a} < 0$   & $\dot{a} + \dot{b} < 0$ & $(\dot{a} + \dot{b}) \dot{a} < 0$ & Profit \\ \hline
    A sell            & B buy             & $\dot{a} < 0$   & $\dot{a} + \dot{b} > 0$ & $(\dot{a} + \dot{b}) \dot{a} > 0$ & Cost \\ \hline
    A buy             & B sell            & $\dot{a} > 0$   & $\dot{a} + \dot{b} < 0$ & $(\dot{a} + \dot{b}) \dot{a} < 0$ & Profit \\ \hline
    \end{tabular}
    \vskip 3pt
    \caption{Impact of Actions on Sign}
    \end{table}

    \section{Two-trader equilibrium strategies}
    \label{sec:two-trader-equilibrium-strategies}

    This section defines what it means for two strategies trading in competition to be in equilibrium. The definition of equilibrium here is analogous to that of Nash equilibrium in standard treatments of game theory. Using our standard setup we start with a unit trader $A$ and a $\lambda$-scaled competitor $B$, with strategies $a(t)$ and $b(t)$ respectively. Write $L_a$ and $L_b$ for the loss functions associated with the total cost of trading $a$ and $b$, as in \cref{eq-loss-fn-a} and \cref{eq-loss-fn-b}.
    
    \begin{defn}[Equilibrium]
    \label{def:equilibrium}
    Let $A$ and $B$ traders with strategies $a(t)$ and $b(t)$. Let $\hat{a}$ be the best-response to $b$; and $\hat{b}$ be the best-response to $a$. Then $a$ and $b$ are {\em in equilibrium} if $\hat{a}=a$ and $\hat{b}=b$. 
    \end{defn}
    
    In simple terms, $A$ and $B$ are in equilibrium if the best-response to $b$ is $a$ and the best-response to $a$ is $b$, and moreover the best-response to the best-response to $b$ is $a$. As it happens this definition is equivalent to the standard definition Nash equilibrium, which we summarize as follows.

    \begin{prop}[Nash equilibrium]
    The traders $A$ and $B$ are in equilibrium  as in Definition \ref{def:equilibrium} if and only if for any unit strategy and $\lambda$-scaled strategy $\tilde{a}\not = a$ and/or $\tilde{b}\not=b$ we have 
    
    \begin{align*}
        \int_0^1 L_{\tilde{a}} \,\text{dt} &> \int_0^1 L_{a} \,\text{dt}, \quad \text{and/or}\\
        \int_0^1 L_{\tilde{b}} \, \text{dt} &> \int_0^1 L_{b} \,\text{dt}
    \end{align*}
    
    \end{prop}

    To find equilibrium strategies we start by deriving equations for joint best-response strategies.

    \subsection{Joint best-response strategies in general}

    Using the same concepts as in \crefrange{sec:best-response-risk-averse}{sec:best-response-eager} we derive equations defining best-response strategies for $A$ and $B$ position-building in competition where $B$ trades a $\lambda$-scaled strategy. First we note that the loss functions for $A$ and $B$ are given as:
    
    \begin{align}
        L(a, b) &= (\dot a + \lambda \dot b) \cdot \dot a + \kappa (a + \lambda b) \cdot \dot a \label{eq-loss-fn-a}\\
        L(b, a) &= (\dot a + \lambda \dot b) \cdot \lambda\dot b + \kappa (a + \lambda b) \cdot \lambda  \dot b \label{eq-loss-fn-b}
    \end{align}
        
    Now applying the Euler-Lagrange equation \cref{eq:euler-lagrange} for $L(a, b)$ and $L(b, a)$ simultaneously yields the {\em equilibrium equations}:
    
    \begin{align}
        \ddot a &= -\frac{\lambda}{2} (\ddot b + \kappa \dot b)     \label{eq:eq-equation-ddota} \\
        \ddot b &= -\frac{1}{2\lambda} (\ddot a + \kappa \dot a)\label{eq:eq-equation-ddotb}
    \end{align}

    with the boundary conditions $a(0)=0, b(0)=0, a(1)=1, b(1)=1$. We will use these in the following sections to analyize and solve for the equilibrium strategies for $a$ and $b$.

    \subsection{Extremal kappa regimes}

    Before studying equilibrium in general we examine two extreme cases in order to build intuition. Fix $\lambda>0$ and. as usual, consider two strategies trading in competition: a unit strategy \(A\) and a \(\lambda\)-scaled strategy \(B\). In equilibrium, the strategies satisfy the equations \crefrange{eq:eq-equation-ddota}{eq:eq-equation-ddotb} above. 
    
    The two extremal cases we investigate are when the contribution to market impact is entirely due to permanent market impact or entirely due to temporary impact. When market impact cost is due entirely to temporary impact, the equilibrium equations reduce to:
    
    \begin{align}
        \ddot a &= -\frac{\lambda}{2} \ddot b  \label{eq:equil-ddota-temp} \\
        \ddot b &= -\frac{1}{2\lambda} \ddot a\label{eq:equil-ddotb-temp}
    \end{align}
    
    again with the boundary conditions $a(0)=0, b(0)=0, a(1)=1, b(1)=1$. Applying the boundary conditions for position-building strategies $a(0), b(0)=0$ and $a(1), b(1)=1$ the solutions are \(a(t) = t\) and \(b(t) = t\). In other words, in the absence of permanent impact, the equilibrium competitive position building reduces to the most trivial of cases. This aligns with intuition: without permanent impact, each trader is exclusively concerned with minimizing total trading costs at each point in time. The best way to achieve this is by spreading trades evenly over the entire duration of trading.
    
    At the other extreme is the cost of trading arises entirely from permanent impact. In this scenario, the equilibrium equations are derived from the following reduced equilibrium equations:
    
    \begin{align*}
        L(a, b) &= \kappa (a + \lambda b) \cdot \dot a \\
        L(b, a) &= \kappa (a + \lambda b) \cdot \lambda  \dot b 
    \end{align*}

    and the Euler-Lagrange equation yields for $L_a$ and $L_b$ respectively: 

    \begin{align*}
        \kappa \dot a + \kappa \lambda \dot b = 0\\
        \kappa \lambda \dot a + \kappa \lambda^2 \dot b = 0\\
    \end{align*}

    dividing through by $\kappa$ in the first equation and $\kappa \lambda$ in the second, these reduce to a single equation

    $$
    \dot a + \lambda \dot b = 0
    $$

    and with the position-building boundary conditions there is no solution. One way to understand what is going on is the look at how the unit trader strategies look at what happens to the solutions $a, b$ for the equilibrium equations \crefrange{eq:eq-equation-ddota}{eq:eq-equation-ddotb} for increasingly large values of $\kappa$, to understanding the limiting behavior of the equations as $\kappa\to \infty$. Here are plots for this scenario:

    \begin{figure}[H]
        \centering
        \includegraphics[width=0.75\linewidth]{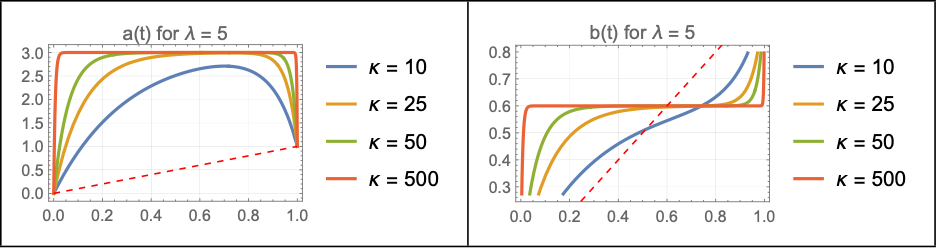}
        \caption{Solutions to the equilibrium equations \crefrange{eq:eq-equation-ddota}{eq:eq-equation-ddotb} as $\kappa$ grows large for a unit trading $A$ and a $\lambda$-scaled trader $B$ trading five times as much as $A$. One can see that the unit trader $a(t)$'s strategy grows toward a strategy of instantly acquiring the three-times its target position instantly and then selling down to target an instant before completion time, while $b(t)$ tends toward acquiring approximately sixty-percent immediately and the rest an instant before completion time.}
        \label{fig:equilibrium-limiting-behavior}
    \end{figure}

    One can see that the limiting strategies in Figure \ref{fig:equilibrium-limiting-behavior} are respectively bucket and barbell shaped (see Section \ref{sec:shapes}). This makes intuitive sense: as market impact becomes dominant, i.e., as $\kappa$ becomes large, the unit strategy $a(t)$ desires to "get ahead" of the pending market impact caused by $b(t)$ so quickly buys up more than its target quantity, hoping to sell at the very end at elevated prices caused by $b(t)$ which is trading five times as much as $a(t)$. A seemingly strong counter to this is for $B$ to buy a portion of its target quantity immediately, imposing elevated prices on $A$ and then delay to the very end completing the position-building.
        
    \subsection{The two-trader equilibrium strategies}
    \label{sec:two-trader-eq-strats}

    Suppose $a$ and $b$ in competition with $b$ $\lambda$-scaled. In this section we derive closed-form expressions for $a$ and $b$ that jointly satisfy \crefrange{eq:eq-equation-ddota}{eq:eq-equation-ddotb}. Solving these yields:

    \begin{align}
        a(t; \kaplam) &= -\frac{\left(1 - e^{-\frac{\kappa  t}{3}}\right) \left(-e^{\kappa /3} \left(e^{\kappa /3}+e^{2 \kappa /3}+1\right) (\lambda +1)+(\lambda -1) e^{\frac{\kappa  t}{3}}+(\lambda -1) e^{\frac{2 \kappa  t}{3}}+(\lambda -1) e^{\kappa  t}\right)}{2 \left(e^{\kappa }-1\right)}  \label{eqs:best-responses-a} \\
        b(t; \kaplam) &= \frac{\left(1 - e^{-\frac{\kappa  t}{3}}\right) \left(e^{\kappa /3} \left(e^{\kappa /3}+e^{2 \kappa/3}+1\right) (\lambda +1)+(\lambda -1) e^{\frac{\kappa  t}{3}}+(\lambda -1) e^{\frac{2 \kappa  t}{3}}+(\lambda -1) e^{\kappa t}\right)}{2 \left(e^{\kappa }-1\right) \lambda }  \label{eqs:best-responses-b}
    \end{align}

    Before moving on to the next section we observe what happens when $\lambda=1$. 

    \begin{equation}
        \label{eq:two-trader-eq-one-uni}
        \begin{split}
        a(t) &= \frac{
                (1- e^{-\frac{\kappa t}{3}}) (e^{\kappa/3} + e^{2\kappa/3} + e^\kappa) 
                }{e^\kappa - 1}
        \end{split}
    \end{equation}
    
    \subsection{Two-trader equilibrium examples}
    \label{sec:two-trader-plots}

\begin{figure}[H]
        \centering
        \includegraphics[width=0.75\linewidth]{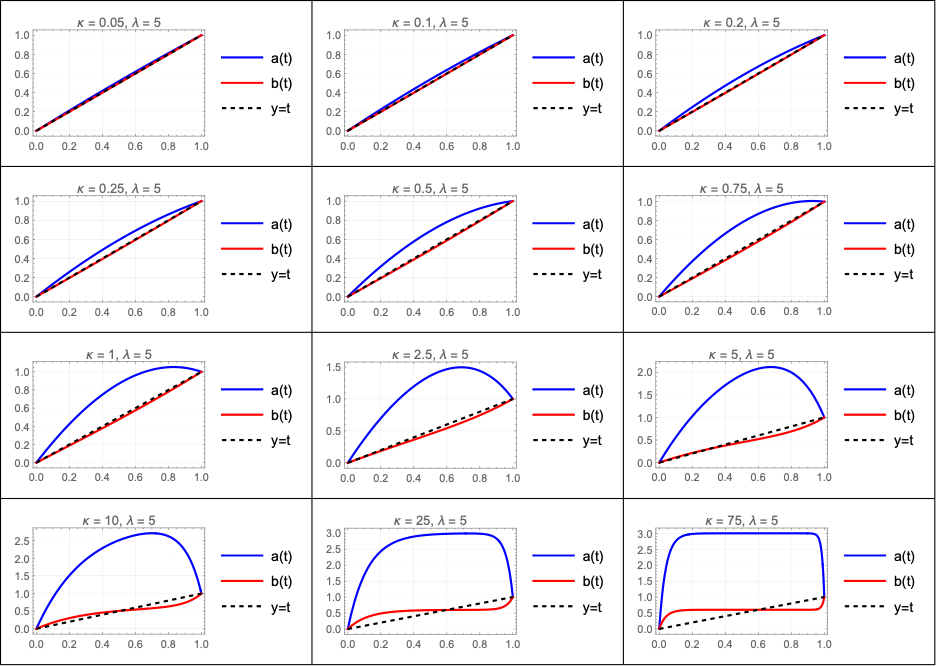}
        \caption{Equilibrium strategies with unit strategy $a(t)$ (in blue) and $\lambda$-scaled strategy $b(t)$ (in red). All of these plots have scaling value $\lambda=5$. The top two rows have $\kappa$-regime (see Section \ref{sec:kappa-regimes}) that are temporary market impact dominated, while the bottom two are permanent market impact dominated.}
        \label{fig:eq-joint-strats}
    \end{figure}

    To conclude our discussion of two-trader equilibrium trading we give some examples of what these strategies look like in general. Figure \ref{fig:eq-joint-strats} shows an example strategy $a(t)$ and $\lambda$-scaled strategy $b(t)$ in equilibrium. The plots show the equilibrium strategies for trader a unit trader $A$ and a $\lambda$-scaled trader $B$ trading in competition. Figure \ref{fig:eq-joint-strats} shows the situation where $\kappa$ is relatively small in comparison to $\lambda$. 
    
    {\bf Note:} The plots for $b(t)$ show the shape of the strategy, without $\lambda$-scaling. Each column of plots shows a single value for permanent impact and each row shows a single values for $\lambda$. The basic message of the plots is that if $A$ and $B$ are trading the same size ($\lambda=1$) then they both simply trade a straight line. As the permanent market impact parameter increases and/or the size that $B$ trades grows, the more convex $A$'s strategy becomes while $B$ remains a straight line. 

\begin{figure}[H]
        \centering
        \includegraphics[width=0.75\linewidth]{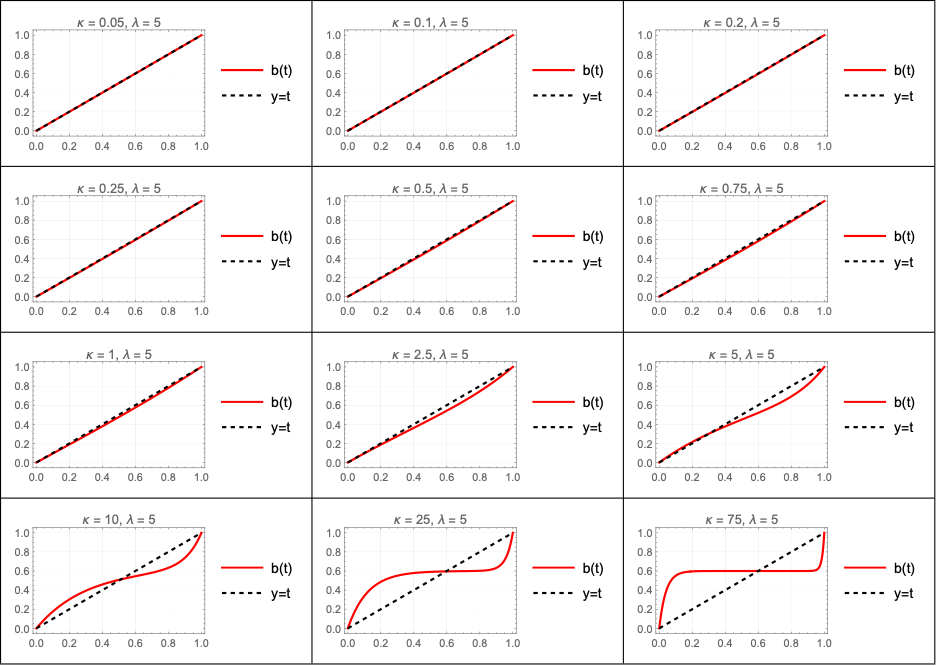}
        \caption{The $\lambda$-scaled strategy from Figure \ref{fig:eq-joint-strats}.}
        \label{fig:bs-strategy-across-kappa-regimes}
    \end{figure}

    Next in Figure \ref{fig:bs-strategy-across-kappa-regimes} we plot just the $\lambda$-scaled strategy from Figure \ref{fig:eq-joint-strats} in order to demonstrate that in sufficiently high $\kappa$, permanent impact dominated regimes, the trading strategy concentrates an increasing proportion of its trading at the start and end of the trading.
    
    A close look at Figures \ref{fig:eq-joint-strats} and \ref{fig:bs-strategy-across-kappa-regimes} suggests the following heuristics for the shapes of the unit and $\lambda$-scaled strategies trading in two-trader competitive equilibrium for different $\kappa$-regimes:

    \begin{itemize}
        \item {\em Negative-$\kappa$, dominated by temporary market impact:} both the unit trader and the $\lambda$-scaled traders trade essentially risk-neutral strategies very close to $y=t$;

        \item {\em Negative-$\kappa$, low-to-moderate relative temporary impact:} the $\lambda$-scaled trader continues to trade an essentially risk-neutral strategies very close to $y=t$; however, the unit trader begins to trade an eager strategy in order to buy ahead of the pending price increases due to the permanent impact to be caused by the $\lambda$-scaled trader;

        \item {\em Positive-$\kappa$, low-to-moderate relative permanent impact:} the unit trader an increasingly eager strategy while the $\lambda$-scaled trader remains more-or-less risk-neutral; and

        \item {\em Positive-$\kappa$, dominated by permanent impact:} now the unit trader begins to trade a bucket strategy and the $\lambda$-scaled trader begins to trade a barbell strategy in response, these shapes as described in Section \ref{sec:shapes}.
    \end{itemize}

\section{Equilibrium when there are many unit traders}
\label{sec:eqilibrium-many-small-traders}

    In the prior section we derived a two-player competitive equilibrium (say, between players $A$, a unit trader, and $B$ and $\lambda$-scaled) and found that the $\lambda$-scaled trader always trades something very close to a risk-neutral strategy while the unit trader trades increasingly eagerly as the size of $\lambda$ grows. We argue that this makes sense because trader $A$ knowing that trader $B$ will push the market up significantly can take advantage of that by purchasing share early, waiting for the price to rise and selling the shares back at a profit. But is this the only possibility?

    Consider the following alternative. Trader $A$ is again a unit trader who wants to acquire one unit of stock between now and time one. They have market surveillance which says that there are one more competitors looking to purchase {\em in aggregate} $\lambda$ units of the stock. The trader considers two alternative possibilities for what is happening:

    \begin{itemize}
        \item {\em One large adversary:} in this alternative there is a single competitor trading $\lambda$ units exactly as in Sections \ref{sec:two-trader-eq-strats} and \ref{sec:two-trader-plots}; and

        \item {\em Many unit trader adversaries:} in this alternative, assuming $\lambda$ is an integer, there are $\lambda$ adversaries each trading a unit strategy. For example, market surveillance says that there will be a total of eleven units traded, one by trader $A$ and the rest traded by ten unit traders.
    \end{itemize}

    Of course there are infinitely many alternatives but closely studying these two will give insights into the central issue at hand. Suppose trader $A$ has no way of identifying which of the two alternatives will transpire. The trader might reason in this situation:

    \begin{quote}
        \em What does it matter whether it is alternative one or alternative two. In either scenario I am trading one unit of the stock and my adversaries are trading ten, so aggregate level of competitive buying I am facing is the same in either scenario. Shouldn't the equilibrium strategy be the same?
    \end{quote}

    This logic, while appealing, is, in fact, incorrect as we shall see next.

    \subsection{The multi-trader symmetric equilibrium strategies}
    \label{sec:many-trader-equilibrium}

    To see why the above logic is incorrect we begin by deriving the equilibrium strategy for $n+1$ traders trading in competition. This is a simple matter of repeating the logic of Section \ref{sec:two-trader-equilibrium-strategies}. 

    If there are $n+1$ unit traders $A_1, \dots, A_{n+1}$ trading in competition we say that $a_i(t)$ is trader $i$'s strategy and has loss function:

    \begin{equation}
        \label{eq:trader-i-loss-fn}
        L_i(t) = \left(\sum_j \dot a_j(t)\right) \dot a_i(t) + 
                      \kappa \cdot  \left(\sum_j  a_j(t)\right) \dot{a}_i(t)
    \end{equation}

    In the above equation the temporary impact from traders $i=1,\dots,n+1$ is $\sum_{i=1}^n \dot a_i(t)$ while the permanent impact is $\kappa\, \sum_{i=1}^n  a_i(t)$. Using the Euler-Lagrange equation \cref{eq:euler-lagrange} applied to trader $i$, that is taking derivatives with respect to $a_i$ and $\dot{a}_i$, we obtain:

    \begin{equation}
        \label{eq:multi-trader-equilibrium-diffeq}
        \ddot a_i = -\frac{1}{2} \left(\sum_{j\ne i} \ddot a_j + \kappa \, \sum_{j\ne i} \dot a_i\right), \quad \text{$a_i(0)=0, a_i(1)=1$ and  $a_i=a_j$ for all $i$}
    \end{equation}

    We will call  \cref{eq:multi-trader-equilibrium-diffeq} the {\em multi-trader symmetric equilibrium equation}. To solve for the {\em symmetric equilibrium} we must solve \ref{eq:multi-trader-equilibrium-diffeq} {\em simultaneously} for $i=1, \dots, n$, assuming that $a_i=a_j$ for all $i, j$. The result is an expression for $a(t)$ in terms of $n$. Then for $n+1$ traders we obtain:

    \begin{equation}
        \label{eq:multi-player-strategy}
        a(t) = \frac{ 1 - e^{-\frac{n \cdot \kappa}{n + 2} \cdot t}}{1 - e^{\frac{n \cdot \kappa}{n + 2}}}        
    \end{equation}

    We call $a(t)$ in \cref{eq:multi-player-strategy} the {\em multi-trader symmetric equilibrium strategy} for $n$ traders. And to clarify the quantities, this means that there are $n$ traders implicitly trading $1$ unit each for a total of $n+1$ units traded. Note that for $n=2$ traders this becomes:

    \begin{equation}
        \label{eq:n-trader-two-traders}
        a(t) = \frac{ e^{\kappa/3} \left( 1 - e^{-\frac{\kappa t}{3}}  \right) }{e^{k/3}-1}
    \end{equation}

    And now we note that 

    \begin{equation}
        \frac{e^{\kappa/3}}{e^{k/3}-1} = \frac{e^{\kappa/3} + e^{2\kappa/3} + e^\kappa }{e^\kappa - 1}
    \end{equation}

    so that \cref{eq:n-trader-two-traders} is equal to:

    \begin{equation}
        a(t) = \frac{
                (1- e^{-\frac{\kappa t}{3}}) (e^{\kappa/3} + e^{2\kappa/3} + e^\kappa) 
                }{e^\kappa - 1}
    \end{equation}

    and this is the same as \cref{eq:two-trader-eq-one-uni}. This shows that the two-trader equilibrium strategy for when each trader trades one unit ($\lambda=1$) is the same as the $n$-trader equilibrium strategy specialized to the case of two-traders.
    
\begin{figure}[h!]
        \centering
        \includegraphics[width=0.75\linewidth]{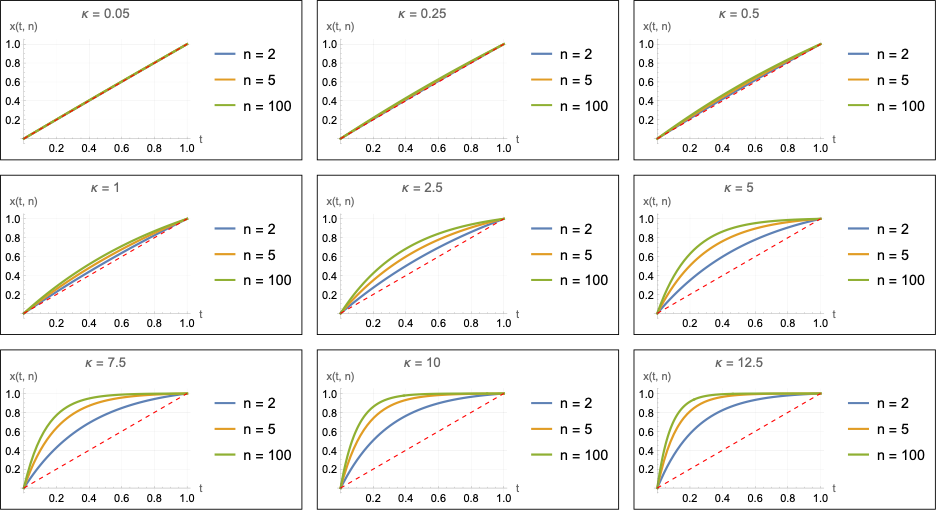}
        \caption{Multi-trader equilibrium \cref{eq:multi-player-strategy} for various values of $\kappa$ and $n$ throughout.}
        \label{fig:multi-trader-equilibrium}
    \end{figure}

    In Figure \ref{fig:multi-trader-equilibrium} we plot these. Each plot shows the multi-trader equilibrium for three values of the number of the total number of traders, labeled $\lambda$ and show values for a total of 10, 50 and 250 traders. As we move down the rows $\kappa$ varies from a low- to high-$\kappa$ regime (see Section \ref{sec:kappa-regimes}). For the low $\kappa$-regime we see the equilibrium strategies are almost risk-neutral, but as $\kappa$ grows the strategies become increasingly eager.

    Next we examine what happens when the number of traders grows extremely large, roughly speaking, modeling the situation in which trader $A$ competes with many traders each trading a single unit. This is equivalent to letting $n\to\infty$ in which case $n\kappa/(n+2) \to \kappa$ and \cref{eq:multi-player-strategy} becomes, in the limit:

    \begin{equation}
        a_i(t) = \frac{e^\kappa - e^{\kappa(1-t)}}{e^\kappa - 1}, \qquad i = 1, \dots, n
    \end{equation}

    and this simplifies to:

    \begin{equation}
        \label{eq:limiting-multi-player-strategy}
        a_i(t) = \frac{1 - e^{-\kappa t}}{1 - e^{-\kappa}}, \qquad i=1, \dots n      
    \end{equation}

\begin{figure}[h!]
        \centering
        \includegraphics[width=0.6666\linewidth]{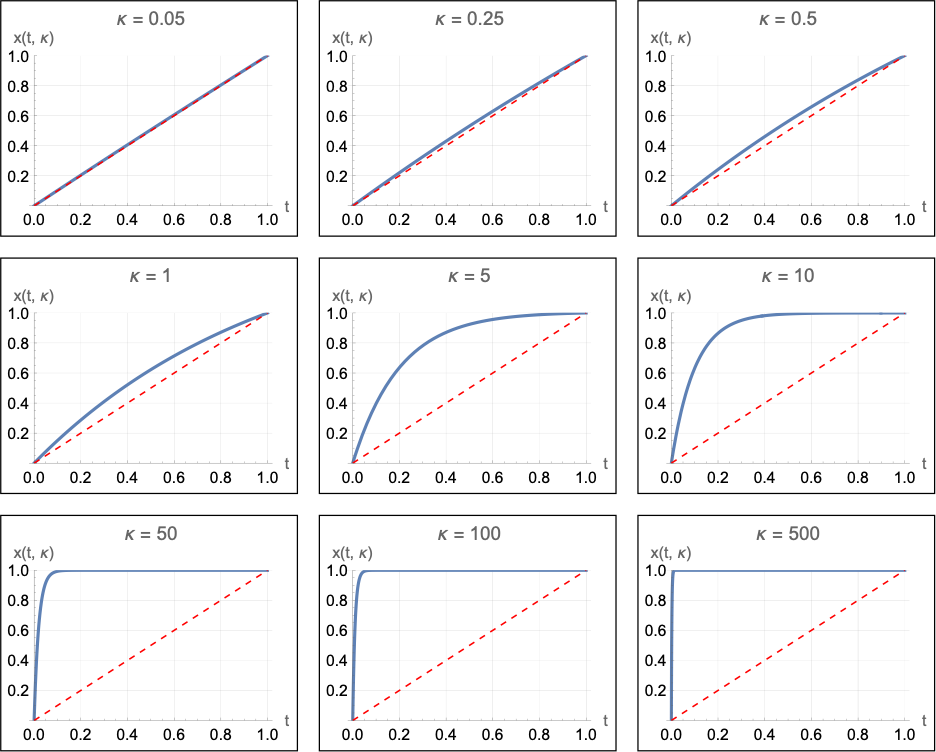}
        \caption{The limiting strategy \cref{eq:limiting-multi-player-strategy} for an essentially infinite number of traders all trading the same strategy in equilibrium. We show the strategies for a variety of $\kappa$ values and a risk-neutral strategy for reference. As $\kappa$ grows large (e.g., the last row) every trader trades increasingly eagerly but never trade a bucket strategy as in Section \ref{sec:shapes}. That is, they never over-buy and re-sell.}
        \label{fig:limiting-multi-player-strategy}
    \end{figure}

    With \cref{eq:limiting-multi-player-strategy} in hand, we plot it with various values of $\kappa$ in Figure \ref{fig:limiting-multi-player-strategy}.

\section{The inverse problem}
\label{sec:inverse-problem}
    
    In this section we develop a method diagnosing the appropriateness of a given strategy intended for trading in competition. As we say in Section \ref{sec:non-equi-best-best-response} if a trader (say, $A$) {\em knows} with certainty that there is a single adversary (say, $B$) trading a known strategy $\lambda\cdot b(t)$ then there is an optimal best-response strategy given by \cref{eq:br-basic}. One obvious problem with this is, {\em what if the adversary $B$ is trading a different strategy $b^\prime$}? 

    \subsection{What is my strategy the best-response to?}

    The first method for examining this issue is to solve the {\em inverse problem}. This amounts to trader $A$ asking {\em if I decide to trader strategy $a(t)$, what strategy is this a best-response for?}. To solve this is a simple matter of applying the Euler-Lagrange equation \cref{eq:euler-lagrange} taking $A$'s strategy as a given to solve for a $\lambda$-scaled strategy $b^*(t)$ in the following loss function, taking $\kappa$ as given:

    \begin{equation}
        \label{eq:inverse-problem-1}
        L(t, a, b^*) = (\dot a + \lambda\dot b^*) \, \dot a + \kappa \,(a + \lambda b^*)\, \dot a
    \end{equation}

    To be clear \cref{eq:inverse-problem-1} is the loss function (that is, the total cost of trading) for $a(t)$ when trading in competition with $\lambda b^*(t)$. We apply the Euler-Lagrange equation and solve for $b^*$:

    \begin{align*}
        \frac{\partial L}{\partial a} &=  2\ddot a + \lambda \ddot b + \kappa \dot a + \kappa \lambda \dot b \\
        \frac{\partial}{\partial t}\frac{\partial L}{\partial \dot a} &= \lambda \dot a
    \end{align*}

    and this yields by the Euler-Lagrange equation:

    $$
    2\ddot a + \lambda \ddot b^* + \kappa \lambda \dot b^* = 0
    $$

    Now, unlike in Section \ref{sec:non-equi-best-best-response} we solve for $\ddot b^*$ which will be the strategy that minimizes the cost for $a(t)$. With this we get:

    \begin{equation}
        \label{eq:inverse-problem-2}
        \ddot b^* = -\frac{1}{\lambda}(2 \ddot a + \kappa \lambda \dot b^*)
    \end{equation}

    with boundary conditions $b(0)=0, b(1)=1$. As a sanity check we can take the solutions to the two-trader equilibrium equations \crefrange{eq:eq-equation-ddota}{eq:eq-equation-ddotb} and the solutions \crefrange{eqs:best-responses-a}{eqs:best-responses-b} and see that if we use the solution for $a(t)$ in the system of equations and then solve for $b^*(t)$ in the inverse problem \cref{eq:inverse-problem-2} we obtain the same $b^(t)=b(t)$ where $b(t)$ is taken from \cref{eqs:best-responses-a}. In some sense this is no surprise as this is precisely the meaning of the equations, however \cref{eq:inverse-problem-2} truly is answers the question {\em if $a(t)$ is $A$'s strategy what is the $b^*(t)$ for which $a(t)$ is the best response?} This has some useful properties which we explain now.

    First suppose that $A$ is trading $a(t)$ and does not know what $B$ is trading but knows the total quantity that will be traded is $\lambda$. Then $A$ may solve for $b^*$ in \cref{eq:inverse-problem-2} and examine it and ask {\em is it sensible that $B$ would trade this way?}

    {\bf Example:} Suppose trader $A$ knows that trader $B$ is going to trader 5 units of stock and $\kappa=2$. If trader $A$ wants to trader a risk-neutral strategy then \cref{eq:inverse-problem-2} states that this is the best-response strategy if $B$ is trading
    
\begin{figure}[H]
        \centering
        \includegraphics[width=0.333\linewidth]{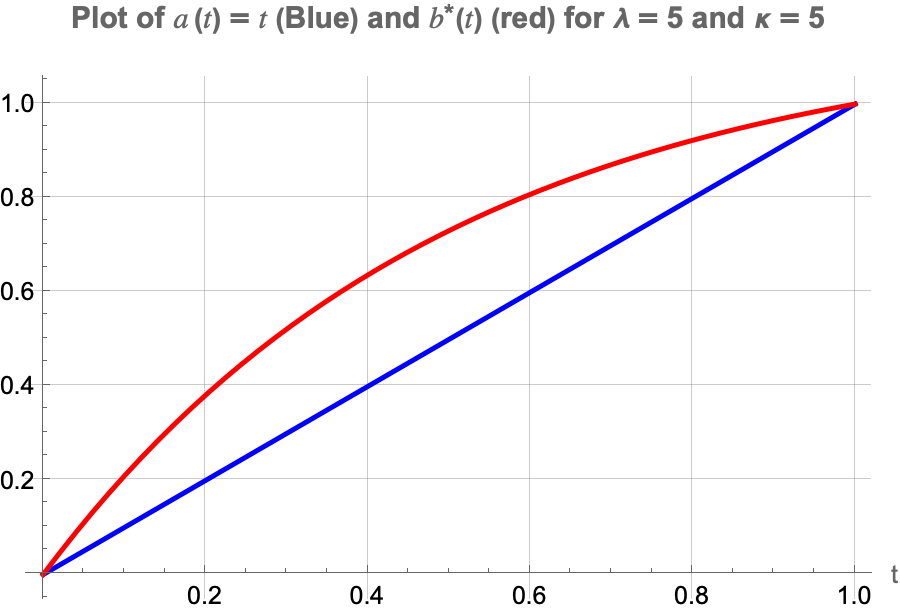}
        \caption{The risk-neutral strategy is the best response to the strategy $b^*(t)$ in this plot.}
        \label{fig:inverse-problem-to-risk-neutral}
    \end{figure}

    We show a plot of \cref{eq:inverse-problem-2} and $a(t)=t$ in Figure \ref{fig:inverse-problem-to-risk-neutral}.

    We also solve the inverse problem when a $\lambda$-scaled trader is trading a strategy $b(t)$ (as usual, its trajectory will be $\lambda \cdot b(t)$). Analogous with above the solution for the strategy $a^*$ which is the best response to the $\lambda$-scaled $b(t)$ is given by

    \begin{equation}
        \label{eq:inverse-problem-for-a}
        \ddot a^* = -(2 \lambda \ddot b + \kappa \dot a^*)
    \end{equation}

    with boundary conditions $a^*(0)=0, a^*(1)=1$. Specifically if we solve \cref{eq:inverse-problem-for-a} using the $\lambda$-scaled solution to the two-trader equilibrium \cref{eqs:best-responses-b}:

    \begin{equation}
        b(t) =-\frac{e^{-\frac{\kappa  t}{3}} \left(e^{\frac{\kappa  t}{3}}-1\right) \left(-e^{\kappa /3} \left(e^{\kappa /3}+e^{2 \kappa /3}+1\right) (\lambda +1)+(\lambda -1) e^{\frac{\kappa  t}{3}}+(\lambda -1) e^{\frac{2 \kappa  t}{3}}+(\lambda -1) e^{\kappa  t}\right)}{2 \left(e^{\kappa }-1\right)}
    \end{equation}

    (note that we solve \cref{eq:inverse-problem-for-a} without using the factor of $\lambda$) we indeed see that it is the best response to 

    \begin{equation}
         a^*(t) = -\frac{e^{-\frac{\kappa  t}{3}} \left(e^{\frac{\kappa  t}{3}}-1\right) \left(-e^{\kappa /3} \left(e^{\kappa /3}+e^{2 \kappa /3}+1\right) (\lambda +1)+(\lambda -1) e^{\frac{\kappa  t}{3}}+(\lambda -1) e^{\frac{2 \kappa  t}{3}}+(\lambda -1) e^{\kappa  t}\right)}{2 \left(e^{\kappa }-1\right)}
    \end{equation}

    which is equivalent to \cref{eqs:best-responses-b}.

    The inverse problem's solution yields another insight. It is an immediate consequence of \cref{eq:inverse-problem-for-a} that if trader $A$ is trading $a(t)$ and this is the best-response to trader $B$'s strategy $\lambda\, b^*$ then if $B$ is trading any other strategy $\lambda \, x(t)$ then 
    
    \begin{equation}
        L(t; a, \lambda\, x) > L(t; a, \lambda\, b^*)
    \end{equation}
        
\section{Strategy selection in the presence of uncertainty}
\label{sec:strategy-selection-uncertainty}

    In this section we look at a numerical analysis of how the framework presented in this paper may be extended for use in the presence of uncertainty. We do not present a complete framework but rather an analysis of a reasonably important special case, in order to illustrate the main features of the problem. Later we present several "hints" as to a more complete framework, but we leave that for a future paper.

    \subsection{Overview}

    We will examine in detail the situation where there are two traders, a unit trader $A$ and a $\lambda$-scaled trader $B$, trading in competition. The difference from the prior sections is that trader $B$ wants to trade an equilibrium strategy but does not know how trader $A$ perceives the market situation. In particular we assume:

    \begin{quote}
        \em Trader $A$ wants to build a position in the same stock as $B$ and is deciding what strategy to trader; $A$ has learned that in addition to its one unit an additional $\lambda$ units will be traded. As such $A$ has good idea of how much will be traded but does not know {\em how} it will be traded. In particular we assume $A$ has narrowed down the possibilities to two: $A$ believes that there is either a single adversary trading the two-trader $\lambda$-scaled equilibrium strategy of Section \ref{sec:two-trader-equilibrium-strategies} or there are $\lambda$ adversaries each trading one unit using the multi-trader symmetric equilibrium strategies of Section \ref{sec:eqilibrium-many-small-traders}.
    \end{quote}
    
    To be clear, the {\em ground truth} is there is only one adversary, however $A$ does not know this.  This section is devoted to analyzing how $B$ will think about strategy selection in light of $A$'s incomplete knowledge. We do this next.

    \subsection{Analysis of trader A's strategy selection}
    \label{sec:trader-a-strategy-selection}
    
    In this, we begin with a trader, $B$, who is trading some strategy $\lambda\cdot b(t)$. Suppose that $B$ believes that it is best to trade an equilibrium strategy, how does $B$ determine its strategy? The issues is that it depends on what $A$ believes is happening. Among all the possibilities, let's suppose that $A$ correctly deduces that in addition to their trading one unit, there will be an additional $\lambda$ units of the stock traded. Further suppose that $A$ boils the situation down to two possibilities, either there is one adversary (the actual situation) or there are $\lambda$ adversaries, each trading one unit. Depending on which of the two possibilities $A$ believes they will trade a different equilibrium strategy and therefore $B$'s best-response will vary as well. To illustrate these two possibilities, we analyze what $B$'s best response is in each case.

    \begin{case}[$B$ believes $A$ has correctly guessed there is only one adversary]
        \label{case:1a}
        If, in fact, $A$ believes there is only one adversary it will trade the strategy \cref{eqs:best-responses-a} which we will identify as $a_{1a}$ and has the form:

    \begin{equation}
          a_{1a}(t) = -\frac{\left(1 - e^{-\frac{\kappa  t}{3}}\right) \left(-e^{\kappa /3} \left(e^{\kappa /3}+e^{2 \kappa /3}+1\right) (\lambda +1)+(\lambda -1) e^{\frac{\kappa  t}{3}}+(\lambda -1) e^{\frac{2 \kappa  t}{3}}+(\lambda -1) e^{\kappa  t}\right)}{2 \left(e^{\kappa }-1\right)}  
    \end{equation}

    and the equilibrium strategy that $B$ should trade given the assumption is that $A$ believes there is a single adversary, and the strategy is given by \cref{eqs:best-responses-b} which we will identify as $b_{1a}$: 

    \begin{equation}
        b_{1a}(t) =\frac{\left(1 - e^{-\frac{\kappa  t}{3}}\right) \left(e^{\kappa /3} \left(e^{\kappa /3}+e^{2 \kappa/3}+1\right) (\lambda +1)+(\lambda -1) e^{\frac{\kappa  t}{3}}+(\lambda -1) e^{\frac{2 \kappa  t}{3}}+(\lambda -1) e^{\kappa t}\right)}{2 \left(e^{\kappa }-1\right) \lambda }  
    \end{equation}

    To be clear, $b_{1a}(t)$ is the strategy that $B$ should trade based on their belief that $A$ guesses that there is one strategy.
    \end{case}
    
    \begin{case}[$B$ believes that $A$ thinks there are a total $\lambda$ adversaries each trading one unit]
    \label{case:1b}
    If, in face, $A$ does incorrectly believe that there are $\lambda$ adversaries, then $A$ will trade \cref{eq:multi-player-strategy} which we identify as $a_{1b}$:

    \begin{equation}
        a_{1b}(t) = \frac{e^{\frac{\lambda \cdot \kappa}{\lambda + 2}} \left(1 - e^{-\frac{\lambda \cdot \kappa}{\lambda + 2} \cdot t}\right)}{e^{\frac{\lambda \cdot \kappa}{\lambda + 2}}-1}       
    \end{equation}

     If $B$ believes $A$ is trading $a_{1b}$ then the best-response, recalling that $B$ will trade $\lambda$ units of the stock, is given by the solutions to the best-response equation \cref{eq:eq-equation-ddotb}, we identify as $b_{1b}(t)$ as follows:

    \begin{equation}
        b_{1b}(t) = \frac{e^{\frac{\kappa  \lambda }{\lambda +2}} \left(\left(\lambda ^2-1\right)
       t+1\right)-e^{-\frac{\kappa  \lambda  (t-1)}{\lambda +2}}+\lambda ^2 (-t)+t}{\lambda ^2
        \left(e^{\frac{\kappa  \lambda }{\lambda +2}}-1\right)}
    \end{equation}        
    \end{case}
    
\begin{figure}[H]
        \centering
        \includegraphics[width=0.75\linewidth]{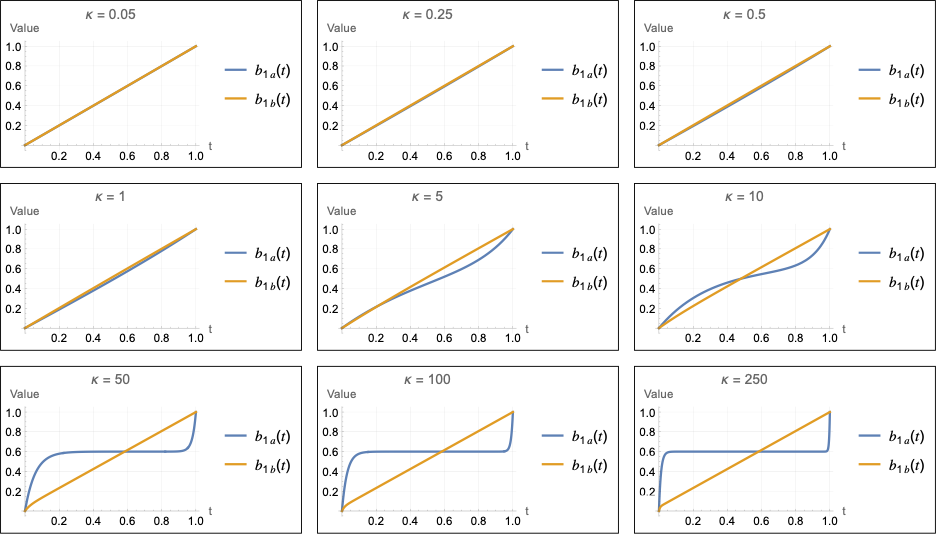}
        \caption{This graph compares the $\lambda$-scaled strategy ($\lambda=5$) when $B$ chooses the best-response according to whether it believes $A$ correctly guesses there is one adversary, $b_{1a}$, or incorrectly guesses there are $\lambda$ adversaries, $b_{1b}.$}
        \label{fig:comparison_b1a_b1b}
    \end{figure}

    Note that strategies $b_{1a}$ and $b_{1b}$ are very similar for small $\kappa$ but for large $\kappa$ they are considerably different. Figure \ref{fig:comparison_b1a_b1b} shows plots comparing the two strategies for $\lambda=5$.
    
    We can see from Figure \ref{fig:comparison_b1a_b1b} that in large $\kappa$-regimes $B$'s strategies vary substantially depending on whether we are in Case 1 or Case 2 above, in which case $B$ trades $b_{1a}$ and $b_{1b}$ respectively. We can now ask the questions, {\em what are the consequences of $B$'s decision versus the ground truth?} We analyze this next.
    
    \subsection{Cost analysis}
    
    Note that in terms of who actually trades what, each of $A$ and $B$ have two possible strategies to choose from $\{a_{1a}, a_{1b} \}$ and $\{b_{1a}, b_{1b} \}$ respectively. This means there are four total possible pairs of strategies that may be traded. Next we plot these four pairs together for a variety of different $\lambda$ and $\kappa$ values.
    
\begin{figure}[h!]
        \centering
        \includegraphics[width=0.65\linewidth]{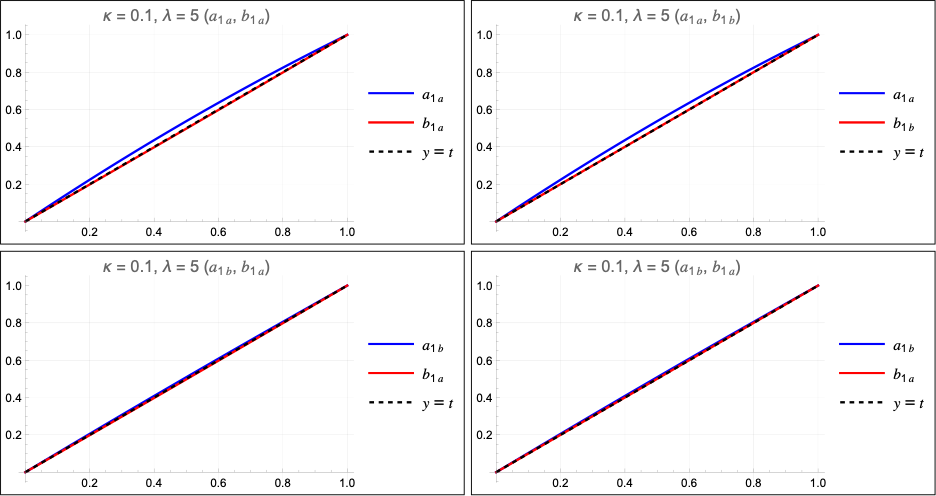}
        \caption{Four possibilities when $\kappa$ is small. In this case all of the strategies concentrate around the risk-neutral strategy and there is no material strategiic uncertainty.}
        \label{fig:error-analysis-kappa-point1-lambda5}
    \end{figure}

    In Figure \ref{fig:error-analysis-kappa-point1-lambda5} we plot the four possible combinations of strategies for $\kappa=0.1$ and $\lambda=5$. We note that in all four cases the strategies are risk-neutral and there will be no cost issues associated with mis-identification.
    
\begin{figure}[h!]
        \centering
        \includegraphics[width=0.65\linewidth]{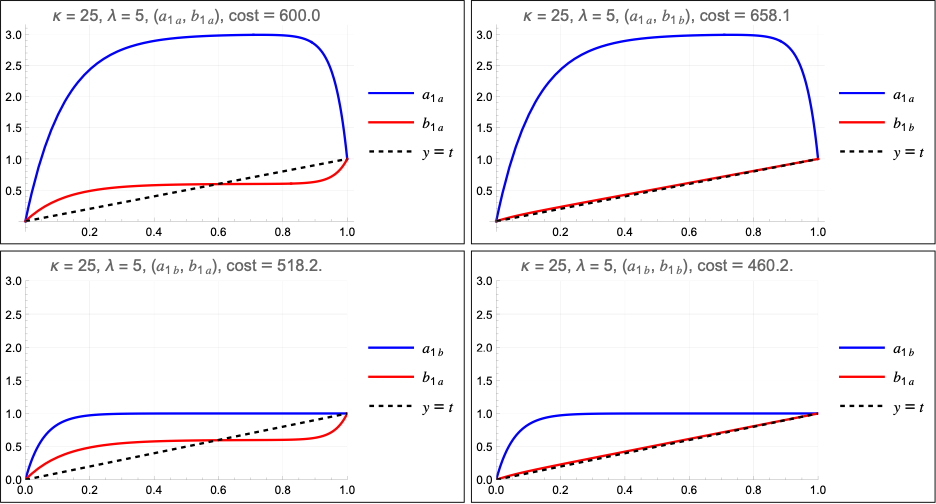}
        \caption{Plots of the four possible combinations of strategies for $A$ and $B$ in analyzing Section \ref{sec:trader-a-strategy-selection} along with the total cost of trading for $B$ in each case. We note that the cheapest strategy combination is when $B$ chooses $b_{1b}$ from Sub-case \ref{case:1b} and assumes that $A$ guesses there are $\lambda$ unit traders, and indeed $A$ trades under this assumption. We investigate these further in the Section \ref{sec:error-analysis}.}
        \label{fig:error-analysis-kappa25-lambda5}
    \end{figure}

    In Figure \ref{fig:error-analysis-kappa25-lambda5} we plot the four possible combinations of strategies for $\kappa=25$ and $\lambda=5$ along the calculated total cost to $B$ of trading the strategy. In this case we see that there is variation in $B$'s strategy and the associated trading costs. We discuss this now. To analyze Figure \ref{fig:error-analysis-kappa25-lambda5} we discuss each plot separately.
    
    We analyze these in more detail in the next section.
    
    \subsection{Error analysis}
    \label{sec:error-analysis}
    
    In this section we analyze the impact of "guessing wrong" in each of the four possible combinations of strategies from Section \ref{sec:strategy-selection-uncertainty}.
    
    \begin{itemize}
        \item $a_{1a}$ and $b_{1a}$: this case means that $B$ is trading {\em assuming} $A$ has correctly guessed there is only one adversary, and $A$ is trading the equilibrium strategy.  Thus the two strategies are equilibrium best-response strategies for one another and $B$ is therefore trading optimally. The total cost of trading for $B$ in this case is 600.0;
    
        \item $a_{1a}$ and $b_{1b}$: In this case $B$ is trading {\em assuming} that $A$ has guessed incorrectly and is trading the equilibrium strategy for $\lambda=5$ unit traders. Put differently, $B$ is trading the best-response for $a_{1b}$, shown in the bottom-left plot while $A$ is actually trading trading $a_{1a}$. The conclusion is that $B$ is trading sub-optimally. The total cost of trading for $B$ in this case is 658.1;
    
        \item $a_{1b}$ and $b_{1a}$: $B$ is again trading assuming $A$ has correctly guessed there is only one adversary, but now $A$ trading $a_{1b}$ the equilibrium strategy assuming there are 25 unit traders and therefore $B$ is also trading sub-optimally. The total cost of trading for $B$ in this case is 518.2; and
    
        \item $a_{1b}$ and $b_{1b}$: Now $B$ is trading under the assumption that $A$ is trading the best-response for 25 adversaries and is again trading optimally. The total cost of trading for $B$ in this case is 460.2.
    \end{itemize}
    
    To summarize the above note that in Figure \ref{fig:error-analysis-kappa25-lambda5} each column represents on of the two possible choices for trader $B$, strategies $b_{1a}$ and $b_{1b}$ respectively. The rows represent the strategy that $A$ has chosen to trade, strategies $a_{1a}$ and $a_{1b}$ and the diagonals of the plot grid are when $B$ is trading the best response and the off-diagonals are when $B$ is not trading the best-response. Note that since $B$ does not know what $A$ is going to guess concerning whether there is one adversaries or $\lambda$ adversaries, then we can regard it as purely a matter of chance whether $B$ ends up trading optimally or not\footnote{The basis for this statement is the $B$ has exhausted all possible market surveillance and is left guessing whether $A$ is going to trade according to their being one or many adversaries}. 
    
    Because of this we can think costs in each column of Figure \ref{fig:error-analysis-kappa25-lambda5} as a random variable with equal probabilities. We summarize these results in Table \ref{tab:matrix_avg_std} and note that the expected value of $B$'s total cost of trading is roughly the same for $b_{1a}$ and $b_{1b}$, however the standard deviation is much lower. 
        
    \begin{table}[H]
        \centering
        \begin{tabular}{@{}lll@{}}
        \toprule
               & $b_{1a}$ & $b_{1b}$ \\ \midrule
        $a_{1a}$ & 600.0   & 658.1   \\
        $a_{1b}$ & 518.1   & 460.2   \\
        \text{avg} & 559.1  & 559.2  \\
        \text{std} & 57.9   & 139.8  \\
        \bottomrule
        \end{tabular}
        \vskip 5pt
        \caption{The total cost of trading for $B$ given the strategies in Figure \ref{fig:error-analysis-kappa25-lambda5} along with the strategy average and standard deviation for $b_{1a}$ and $b_{1b}$.}
        \label{tab:matrix_avg_std}
    \end{table}

    Examining Figure \ref{fig:error-analysis-kappa25-lambda5} we are reminded that $B$ is facing one of two strategies whether it chooses $b_{1a}$ or $b_{1b}$, and we see that in a certain sense because the strategies have roughly the expected cost, if $B$ wishes to minimize variance, then it is best to choose strategy $b_{1a}$.   
        
\section{Strategy selection}
    \label{sec:strategy-selection}

    Section \ref{sec:strategy-selection-uncertainty} presents a crude means of analyzing selection criteria in the case where $B$ wants to select an equilibrium strategy but is uncertain of how $A$ will proceed. We are able to nevertheless select a strategy by viewing the set of possible strategies that $A$ would select as determining the set of strategies $B$ would select. In this way we were able to compute a rudimentary expected value for each possible equilibrium strategy that $B$ can select. This suggests a more rigorous notion, which we explain here.

    \subsection{Probabilistic strategy selection (speculative)}
    \label{sec:probabilistic-strategy-selection}

    In this section we analyze the case where $A$ is a trader wishing to buy one unit of stock $S$ between time $t=0$ and $t=1$, and has collected data on what strategies may be trading in competition. We start by defining some terms.

    \subsection{Notation and terms}
    \label{sec:prob-selection-notation-terms}
    
    The set $\cal S$ is defined as the set of all $\lambda$-scaled strategies $b(t)$ that $B$ could possibly be trading, and write $b_\lambda$ for the corresponding $\lambda$-scaled strategy which arises from the two-trader equilibrium strategy \cref{eqs:best-responses-b} scaled by $\lambda$. For sure a $b_\lambda$ write $a_\lambda(b)$ for $A$'s best-response strategy to $b_\lambda$ and $\dot a_\lambda(b)$ for its time derivative. With this said we can express the cost of trading $a_\lambda$ in competition with $b_\lambda$ as  

    \begin{equation}
    \label{eq:total-expected-cost-trading}
        C(a_\lambda, b_\lambda; \kappa) = \int_0^1 (\dot a_\lambda + \lambda\dot b_\lambda)\dot a_\lambda + \kappa (a_\lambda + \lambda b_\lambda)\dot a_\lambda  \,\, \text{dt}     
    \end{equation}

    the cost of trading the best-response strategy to the $\lambda$-scaled strategy $b$ (for $\lambda b \in \cal S$) assuming market impact parameter $\kappa$. Now assume we can equip the set $\cal S$ with a probability measure $m$ then we can express the {\em expected cost of trading} with respect to the measure $m$ as: 

    \begin{align*}
    E[C | {\cal S}] &= \int_{b_\lambda \in \cal S} C(a_\lambda, b_\lambda; \kappa) \,\,\text{dm} \\
                    &= \int_{b_\lambda \in \cal S} \int_0^1 (\dot a_\lambda + \lambda\dot b_\lambda)\dot a_\lambda + \kappa (a_\lambda + \lambda b_\lambda)\dot a_\lambda  \, \text{dt} \,\,\text{dm}
    \end{align*}
    
    from which we may define its corresponding variance:

    $$
    \text{Var}[C| {\cal S}] = E[C^2 | {\cal S}] - E[C | {\cal S}]^2
    $$

    and proceed with a mean-variance analysis similar to as in Section \ref{sec:strategy-selection-uncertainty}. 

    \subsection{Example computation with a log-normal distribution}

    Using the setup in Section \ref{sec:prob-selection-notation-terms} we define $\cal S$ as the set of $\lambda$-scaled two-trader equilibrium strategies trading in competition with $A$. Assume that the market impact parameter $\kappa$ is known\footnote{We can repeat these computations making $\kappa$ probabilistic just as easily.}. In particular suppose that we equip $\mathbf{R}^+$ with the log-normal probability density for some mean $\mu>0$ and standard deviation $\sigma>0$. Then the density is given by
    
    $$
    f(x; \mu, \sigma) = \frac{1}{x \sigma \sqrt{2 \pi}} \exp\left(-\frac{(\ln x - \mu)^2}{2 \sigma^2}\right), \quad x > 0, \, \sigma > 0
    $$
    
    and write $\text{dm}$ for the associated measure on $\mathbf{R}^+$. Then write $b_\lambda(t)$ for the two-trader $\lambda$-scaled equilibrium strategy of Section \ref{sec:two-trader-eq-strats}. Then the expected total cost of trading \cref{eq:total-expected-cost-trading} becomes:

    \begin{align}
    E[C|b_\lambda, \lambda\in \mathbf{R}^+] &= \int_{\lambda\in \mathbf{R}^+} C(a_\lambda, b_\lambda; \kappa) \,\, \text{dm}    \label{eq:expected-value-log-normal-equi} \\
                &= \int_{\lambda\in \mathbf{R}^+} \int_0^1 (\dot a_\lambda + \lambda\dot b_\lambda)\dot a_\lambda + \kappa (a_\lambda + \lambda b_\lambda)\dot a_\lambda  \,\, \text{dt} \,\, \text{dm}\nonumber
    \end{align}
    
    We leave the computation of the expected cost \cref{eq:expected-value-log-normal-equi} and its associated variance for a future paper.

    \section{Parameter mis-estimation}
    \label{sec:mis-estimation}

\begin{figure}[h!]
        \centering
        \includegraphics[width=0.5\linewidth]{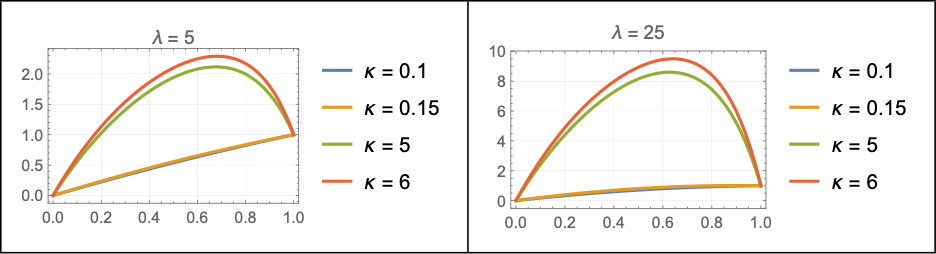}
        \caption{For the two-trader equilibrium strategies of Section \ref{sec:two-trader-eq-strats} this is a plot depicting what happens to a unit trader's strategy when $\kappa$ or $\lambda$ is mis-estimated. Each plot shows slightly different values for $\kappa$ for two levels of $\kappa$. The left plot is for $\lambda=5$ and the right for $\lambda$. We see that for low kappa regimes, strategies do not differ substantially. In high kappa regimes the story is quite different. Changes in $\lambda$ and $\kappa$ can result in quite different strategies, but what is the impact on cost.}
        \label{fig:mis-estimation-kappa}
    \end{figure}

    In this section we briefly touch upon what happens when the parameters related to optimal position-building. Consider what happens when we mis-estimate the value of $\kappa$ by examining Figure \ref{fig:mis-estimation-kappa}. As with Section \ref{sec:strategy-selection} we do not present a complete picture but rather potential directions for future development.

    \subsection{Costs arising from errors in the market impact coefficient}

    We start by giving a numerical exposition of how changes in the market impact coefficient $\kappa$ effect the total cost of trading in two-trader equilibrium strategies, specifically for the unit strategy $a(t)$.

    Another view of how changes in $\kappa$ impact strategies is given in Figure in which we explicitly plot the two-trader equilibrium strategies $a(t), b(t)$ (see \crefrange{eqs:best-responses-a}{eqs:best-responses-a}) for various levels of $\kappa$. Each plot shows $a(t)$ for values of $\kappa$ shifted by 25\% along with $b(t)$ in the right panel and the difference between the two $a(t)$ strategies in the left panel.

    \begin{figure}[h!]
        \centering
        \includegraphics[width=0.5\linewidth]{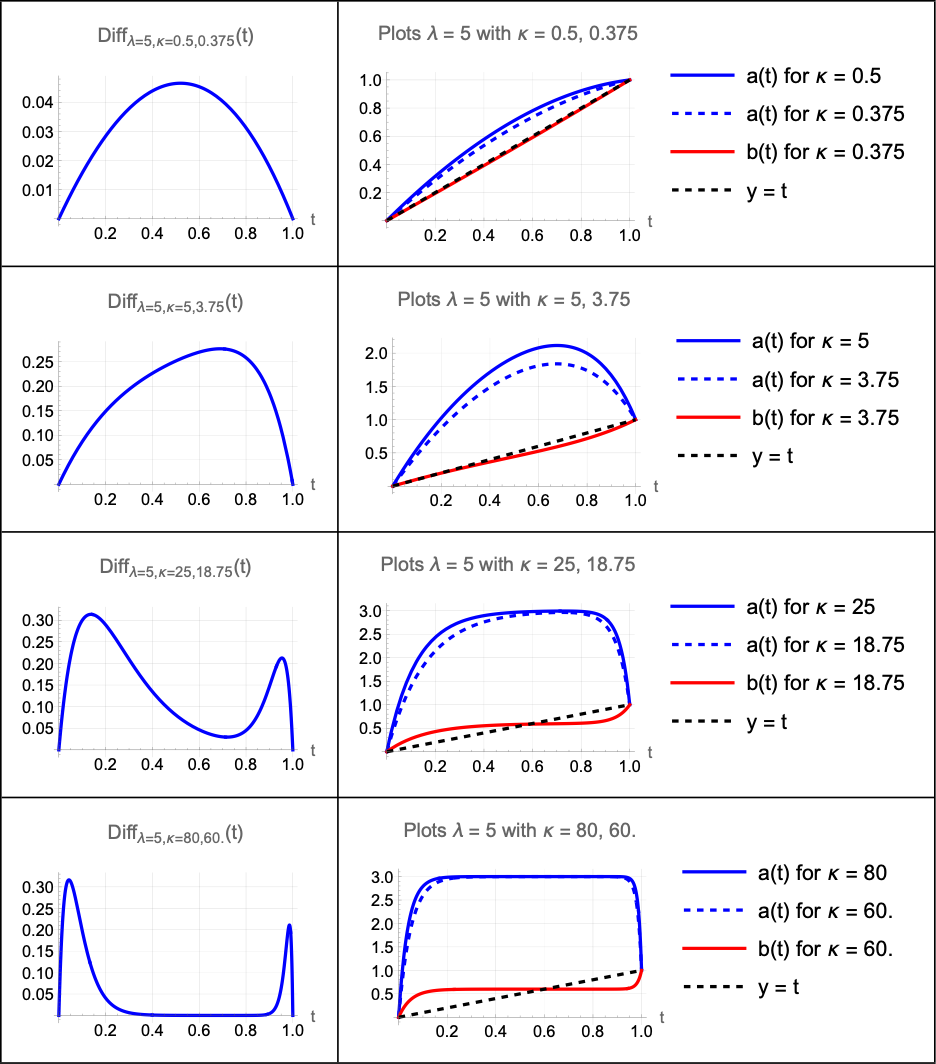}
        \caption{A visual demonstration of the impact of changes in $\kappa$ on two-trader equilibrium strategies.}
        \label{fig:equilibrium-kappa-diffs}
    \end{figure}

    Examining Figure \ref{fig:equilibrium-kappa-diffs} the question that arises is what are the cost differences associated with the strategies. To understand this we compute the impact costs of mis-estimating $\kappa$ in Table \ref{tab:impact-results}. The table shows pairs of strategies, each the unit trader strategy $a(t)$ in \cref{eqs:best-responses-a} for related values of $\kappa$.

    \begin{table}[h!]
    \centering
    \begin{tabular}{|c|c|c|c|c|c|}
    \hline
    $\lambda$ & $\kappa$ & Temp Impact & Perm Impact & Total Impact & Total Diff \\
    \hline
    5 & 0.1 & 2.01 & 0.10 & 2.10 &  \\
    5 & 0.075 & 2.0 & 0.07 & 2.08 & -1.28\% \\
    \hline
    5 & 0.5 & 2.17 & 0.43 & 2.61 &  \\
    5 & 0.375 & 2.10 & 0.34 & 2.44 & -6.60\% \\
    \hline
    5 & 1 & 2.69 & 0.74 & 3.43 &  \\
    5 & 0.75 & 2.39 & 0.60 & 2.99 & -12.68\% \\
    \hline
    5 & 5 & 14.52 & 0.28 & 14.79 &  \\
    5 & 3.75 & 9.97 & 0.73 & 10.70 & -27.68\% \\
    \hline
    5 & 25 & 84.99 & -4.99 & 79.99 &  \\
    5 & 18.75 & 63.66 & -3.71 & 59.96 & -25.05\% \\
    \hline
    5 & 100 & 340.00 & -20.00 & 320.00 &  \\
    5 & 75 & 255.00 & -15.00 & 240.00 & -25.00\% \\
    \hline
    \hline
    \end{tabular}
    \vskip 5pt
    \caption{Pairs of market impact costs for the unit strategy $a(t)$, in two-trader equilibrium strategies for $\lambda=5$ and various $\kappa$ values. Each pair of rows shows the temporary, permanent and total market impact costs for strategy $a(t)$ that differ only in a 25\% shift in the value of $\kappa$. We see that as the level of $\kappa$ itself grows, the cost sensitivity grows as well.}
    \label{tab:impact-results}
    \end{table}

    As a check on the impact of the level of $\lambda$ on the sensitivity to $\kappa$ we re-compute Table \ref{tab:impact-results} for $\lambda=25$ and present the results in Table \ref{tab:impact-results-1}.

    \begin{table}[h!]
    \centering
    \begin{tabular}{|c|c|c|c|c|c|}
    \hline
    $\lambda$ & $\kappa$ & Temp Impact & Perm Impact & Total Impact & Total Diff \\
    \hline
    25 & 0.1 & 2.22 & 0.09 & 2.30 &  \\
    25 & 0.075 & 2.12 & 0.07 & 2.19 & -4.96\% \\
    \hline
    25 & 0.5 & 7.43 & 0.15 & 7.58 &  \\
    25 & 0.375 & 5.06 & 0.18 & 5.24 & -30.90\% \\
    \hline
    25 & 1 & 23.42 & -0.36 & 23.06 &  \\
    25 & 0.75 & 14.14 & -0.02 & 14.12 & -38.75\% \\
    \hline
    25 & 5 & 388.63 & -19.57 & 369.06 &  \\
    25 & 3.75 & 248.94 & -11.94 & 236.99 & -35.78\% \\
    \hline
    25 & 25 & 2459.70 & -130.96 & 2328.80 &  \\
    25 & 18.75 & 1841.70 & -98.02 & 1743.70 & -25.12\% \\
    \hline
    25 & 100 & 9841.30 & -524.0 & 9317.30 &  \\
    25 & 75 & 7381.00 & -393.00 & 6988.00 & -25.00\% \\
    \hline
    \hline
    \end{tabular}
    \vskip 5pt
    \caption{A companion to Table \ref{tab:impact-results}, but with $\lambda=25$. We note that as $\kappa$ grows large the sensitivity to 25\% changes in $\kappa$ levels off at 25\%, the same as the case with $\kappa=5$, however in this case the sensitivity to errors in $\kappa$ is much larger for moderate levels of $\kappa$, for instance at $\kappa=0.5$ when $\lambda=5$ the sensitivity is 6.6\% whereas when $\lambda=25$ the sensitivity for the same level of $\kappa$ is 31\%.}
    \label{tab:impact-results-1}
    \end{table}

    Next we discuss ways to calculate mis-estimation risk more analytically. 

    \subsection{Analytic evaluation of mis-estimation cost}

    Consider two traders $A$ and $B$ trading strategies $a(t)$ and $\lambda b(t)$. For this discussion it is not relevant how these two strategies are determined, but for example they can arise as two-trader equilibrium strategies. Clearly the specific strategies were arrived at with assumptions about $\lambda$ and the market impact parameter $\kappa$. The total cost of trading formulas for $A$ and $B$ trading in competition are given as in Section \ref{sec:total-cost-of-trading-in-comp} and we re-write these here showing the explicit dependence on $\lambda$ and $\kappa$:

    \begin{align}
        C(a; b, \lambda, \kappa) = \int_0^1 \big(
            \dot a(t; \lambda, \kappa) + \lambda \dot b(t; \lambda, \kappa) \big) \dot a(t; \lambda, \kappa) +
            \big(
                a(t; \lambda, \kappa) + \lambda b(t; \lambda, \kappa) \big) \dot a(t; \lambda, \kappa) \,\,\text{dt} \label{eq:explicit-total-cost-a-with-b}\\
        C(b; a, \lambda, \kappa) = \int_0^1 \big(
            \dot a(t; \lambda, \kappa) + \lambda \dot b(t; \lambda, \kappa) \big) \lambda \dot b(t; \lambda, \kappa) +
            \big(
                a(t; \lambda, \kappa) + \lambda b(t; \lambda, \kappa) \big) \lambda\dot b(t; \lambda, \kappa) \,\,\text{dt}\label{eq:explicit-total-cost-b-with-a}
    \end{align} 

    The function $C(a; b, \lambda, \kappa$ is the total cost of strategy $a$'s trading while in competition with $\lambda \, b$, with the dependence on $\kappa$ and $\lambda$ made explicit. Similarly, $C(b; a, \lambda, \kappa)$ is the total cost of $b$'s trading while in competition with $a$. With this we can now for the partial derivatives:

    \begin{align}
        \frac{\partial C(a; b, \lambda, \kappa)}{\partial \lambda} &= \text{Sensitivity of $a$'s total cost of trading to $\lambda$}\\
        \frac{\partial C(a; b, \lambda, \kappa)}{\partial \kappa} &= \text{Sensitivity of $a$'s total cost of trading to $\kappa$}\\
        \frac{\partial C(b; a, \lambda, \kappa)}{\partial \kappa} &= \text{Sensitivity of $b$'s total cost of tradiyng to $\kappa$}\\
        \frac{\partial C(b; a, \lambda, \kappa)}{\partial \kappa} &= \text{Sensitivity of $b$'s total cost of trading to $\lambda$}
    \end{align}

    We call the above functions the {\em cost sensitivity} functions. The above partial derivatives may be computed either analytically or numerically and represent the sensitivity of the total cost of trading to changes in $\lambda$ or $\kappa$ and suggest they may be used to guide strategy selection using the dictum {\em when in doubt, choose the less sensitive strategy.}
          
    \section{Two-trader equilibrium with risk aversion}
    \label{sec:two-trader-equi-with-risk-aversion}

    Holding risk in this context refers to a general aversion to holding the stock during the course of building the position. This may arise, for example, because while the predominant possibility is that the stock will rise during the position-building process and then "jump" when the catalyst occurs, there is a small chance that some bad news will occur and cause the price to fall. In this section we augment the two-trader equilibrium strategies discussed in Section \ref{sec:two-trader-equilibrium-strategies} by adding a term the penalizes aversion to holding risk, as discussed in Section \ref{sec:almgren-chriss-review} and which is a key feature of optimal execution models as in \cite{almgren1997optimal} and \cite{almgren2001optimal}. 
    
    \subsection{Augmented equilibrium equations}
    
    To add risk-aversion to two-trader equilibrium we augment the loss functions \cref{eq-loss-fn-a}{eq-loss-fn-b} with a term proportional to $\sigma^2$ times the holding itself, where $\sigma$ is the volatility of the stock over the holding period:

    \begin{align}
        L_a &= (\dot a + \lambda \dot b) \, \dot a + \kappa (a + \lambda b) \, \dot a + \xi_a \, \sigma^2\, a^2 \label{eq-loss-fn-a-with-risk}\\
        L_b &= (\dot a + \lambda \dot b) \, \lambda\dot b + \kappa (a + \lambda b) \, \lambda  \dot b + \xi_b \, \sigma^2 \,b^2\label{eq-loss-fn-b-with-risk}
    \end{align}

    Note that these equations are the same as \cref{eq-loss-fn-a}{eq-loss-fn-b} but they penalize the total volatility a trader's position holds during the course of building the position.

    Using the Euler-Lagrange equation \cref{eq:euler-lagrange} we then obtain the system of differential equations 

    \begin{align}
        2\ddot a + \lambda \ddot b + \kappa\lambda \dot b  - 2 \xi_a \sigma^2 a &= 0\\
        \ddot a + 2 \lambda \ddot b + \kappa\dot a - 2 \frac{\xi_b}{\lambda} \sigma^2 b &= 0  
    \end{align}

    These in turn yield the {\em two-trader equilibrium equations with risk} analogous to \crefrange{eq:eq-equation-ddota}{eq:eq-equation-ddotb}:

    \begin{align}
        \ddot a &= -\frac{\lambda}{2} (\ddot b + \kappa \dot b) +  \xi_a \sigma^2 a \label{eq:ddota-two-trader-with-risk} \\
        \ddot b &= -\frac{1}{2\lambda} (\ddot a + \kappa \dot a) +\frac{\xi_b}{\lambda^2} \sigma^2 b \label{eq:ddotb-two-trader-with-risk}
    \end{align}

    with, as usual, the boundary conditions $a(0)=0, b(0)=0, a(1)=1, b(1)=1$. Note that in these equations $\xi_a$ and $\xi_b$ are trader-specific trade-off parameters that add risk-aversion to cost-of-trading, analogous to \cref{eq:loss-function-alm-chr}.

    \subsection{Exploration of two-trader equilibrium with risk-aversion}

\begin{figure}[h!]
        \centering
        \includegraphics[width=0.666\linewidth]{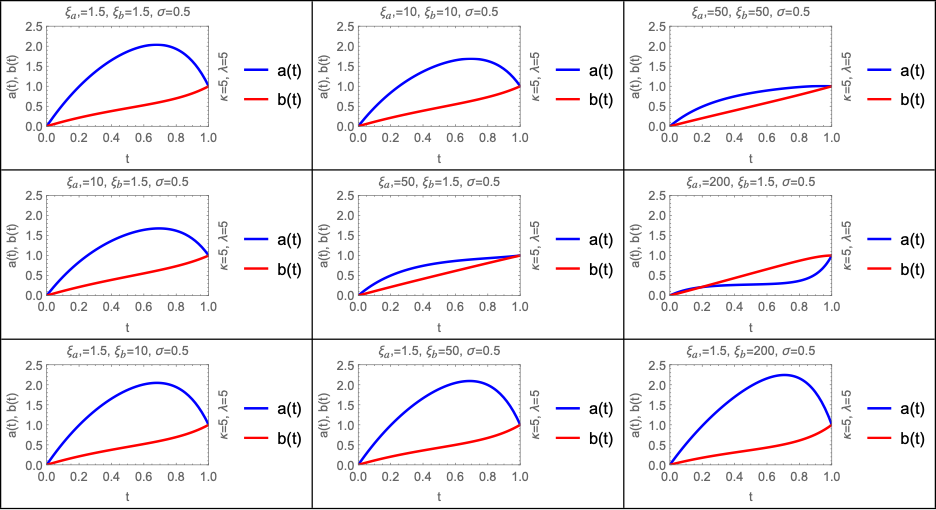}
        \caption{Two-trader equilibrium strategies traders $A$ and $B$ trading a unit and $\lambda$-scaled strategy, respectively. In these plots $\lambda=5$ and $\kappa=5$ (the market impact coefficient) and $\sigma=0.5$ throughout. These plots show three different scenarios for the relationship between trader $A$'s and $B$'s level of risk-aversion. The first row shows three levels of risk aversion. In the first row $\xi_a = \xi_b$, in the second $\xi_a>\xi_b$ and the third $\xi_a<\xi_b$. From left to right the absolute level of risk aversion grows in each row and we the $y$-axis, the level of stock purchase has the same scale in each plot.}
        \label{fig:equi-with-risk-1}
    \end{figure}

    We briefly provide some numerical examples of the trading strategies satisfying \crefrange{eq:ddota-two-trader-with-risk}{eq:ddotb-two-trader-with-risk}. We refer to the solutions as the {two-trader equilibrium strategies with risk-aversion} for the remainder\footnote{We do not present the equations here but see Appendix \ref{sec:appendix} for that Mathematica code that produced Figure \ref{fig:equi-with-risk-1}.}. In Figure \ref{fig:equi-with-risk-1} we plot the solutions to these equations for various levels of risk aversion and constant levels of $\lambda$ and $\kappa$ (see the caption for details). The plots show the expected tension that arises between risk-aversion and the desire to minimize total cost of trading. 

    We draw the following conclusions examining Figure \ref{fig:equi-with-risk-1} for the case where $B$ (the red line) has $\lambda=5$ and $A$ (the blue line) is a unit trader and $\sigma=0.5, \kappa=5$ throughout:

    \begin{itemize}
        \item When $A$ and $B$ have the same level of risk-aversion (the first row in Figure \ref{fig:equi-with-risk-1}, $A$ nevertheless trades an {\em eager} strategy;

        \item When $A$ has {\em greater} risk aversion than $B$ (the second row), $A$ trades an eager strategy until the absolute level of risk aversion grows very high (that last plot in the second row); and

        \item When $B$ has {\em greater} risk aversion than $A$ (the second row), $B$ trades a roughly risk-neutral strategy and $B$ trades a very eager strategy.
    \end{itemize}

    With these conclusions in mind we repeat the plots of Figure \ref{fig:equi-with-risk-1} in Figure \ref{fig:equi-with-risk-2} but with $\sigma=2.0$, four times that of the Figure \ref{fig:equi-with-risk-1}:

    \begin{figure}[H]
        \centering
        \includegraphics[width=0.6666\linewidth]{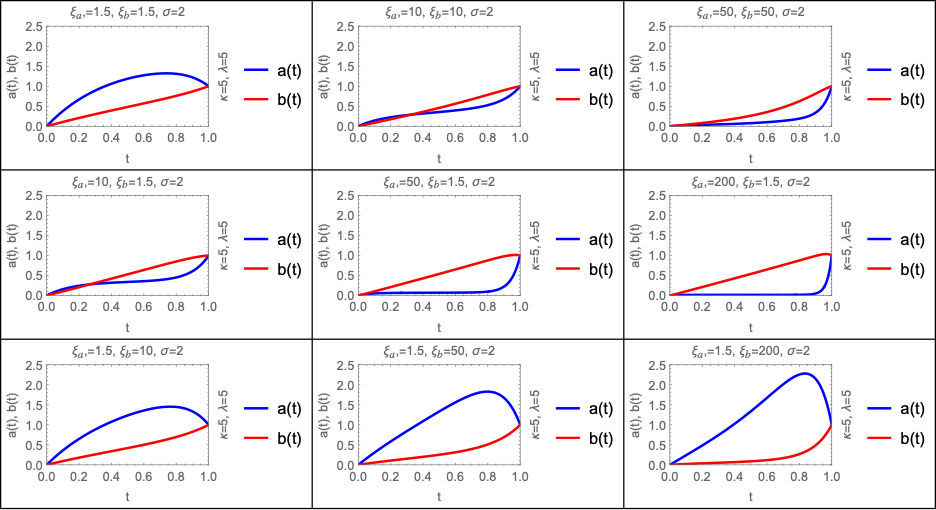}
        \caption{A repeat of Figure \ref{fig:equi-with-risk-1} but with $\sigma=2.0$. The values for $\xi_a, \xi_b, \lambda, \kappa$ are the same as in the prior figure. We see when the absolute level of risk grows, the unit trading strategy $a(t)$ can become risk averse; however, in the bottom row where trader $B$ is more risk-averse than $A$, $B$ is forced to trade an eager strategy despite its own risk-aversion.}
        \label{fig:equi-with-risk-2}
    \end{figure}

    We see in Figure \ref{fig:equi-with-risk-2} that increasing the volatility of the stock by a factor of four has several notable changes as compared to the prior Figure \ref{fig:equi-with-risk-1}:

    \begin{itemize}
        \item When trader $A$'s level of risk-aversion (as given by $\xi_a$) is moderately high (as opposed to very high), the strategy becomes risk-averse as is seen from the center plot and the last column of the first two rows in the two figures; and

        \item Trader $A$ does not engage in significant overbuying unless $B$ is significantly more risk-averse than $A$;        
    \end{itemize}

    Next in Figure \ref{fig:equi-with-risk-3} we plot the same scenarios as Figure \ref{fig:equi-with-risk-1} but this time with a much lower level of permanent market impact relative to temporary, setting $\kappa=0.1$ throughout:

    \begin{figure}[h!]
        \centering
        \includegraphics[width=0.666\linewidth]{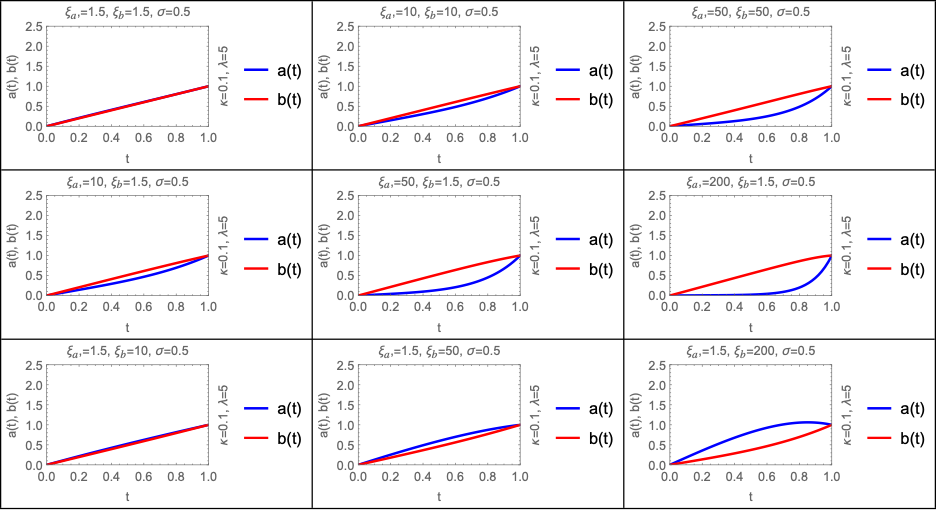}
        \caption{The same plots as in Figure \ref{fig:equi-with-risk-1} but with much {\em lower} levels of permanent market impact relative to temporary ($\kappa=0.5$, throughout). With the concern of "paying up" for the stock toward the end of position-building, $A$ trades a risk-averse strategy in most scenarios, though when $B$'s level of risk-aversion far exceeds $A$'s (bottom row, third column) $A$ reverts to trading an eager strategy.} 
        \label{fig:equi-with-risk-3}
    \end{figure}

    Figure \ref{fig:equi-with-risk-3} shows a similar pattern to \ref{fig:equi-with-risk-3} but much more muted. The conclusion in this scenario is that the importance of permanent impact is significantly reduced and so trader $A$ can afford to trade a risk-averse strategy without paying excessive prices in the latter portion of the position-building.

\begin{figure}[H]
        \centering
        \includegraphics[width=0.6666\linewidth]{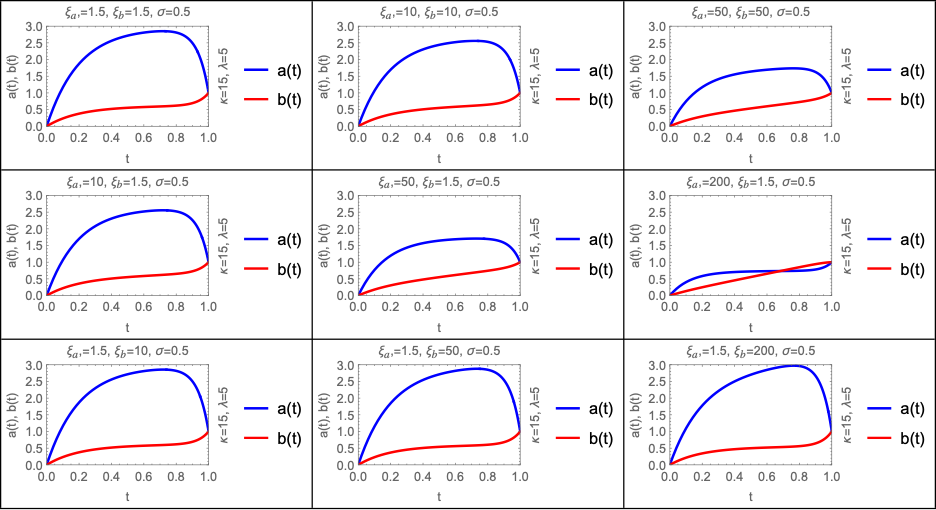}
        \caption{The same plots as in Figure \ref{fig:equi-with-risk-1} but with much {\em higher} levels of permanent market impact relative to temporary. The possibility of the stock price being significantly bid up, we see that outside of very high levels of risk-aversion, trader $A$ continues to trade an eager strategy.}
        \label{fig:equi-with-risk-4}
    \end{figure}

    As a final demonstration we plot the same as Figure \ref{fig:equi-with-risk-1} but this time with $\kappa=15$ and $\lambda=5$. In these plots we see that the pressure of "buying ahead" of trader $B$ (who is trading five times as much as $A$) generally places $A$ in the eager trading regime. The sole exception is when $A$'s risk aversion is extremely high relative to $B$'s, as in the second row and far-right column of Figure \ref{fig:equi-with-risk-4}.

\section{Summary}
    
    In this paper, we developed a framework for understanding and optimizing position-building strategies in competitive trading scenarios. Our analysis is grounded in a game-theoretic setting, where each trader's actions are represented by their trading strategies, and the primary objective is to minimize a cost function that accounts for both temporary and permanent market impact. We introduced and analyzed the concept of equilibrium strategies, where each trader's strategy is the best possible response to the other trader's strategy.
    
    The framework was explored through a detailed study of various strategy types, including risk-neutral, risk-averse, and eager strategies. We also introduced more complex trading patterns such as bucket and barbell strategies, which traders may adopt under specific market conditions. 
    
    Through the use of differential equations and the Euler-Lagrange equation, we derived optimal strategies for different competitive scenarios. We further extended the analysis to multi-trader environments, showing how strategies evolve as the number of competing traders increases.
    
    A key aspect of the paper is the exploration of the impact of the market impact coefficient, \(\kappa\), on the optimal strategies. We identified different \(\kappa\) regimes—temporary impact-dominated and permanent impact-dominated—and demonstrated how these regimes influence the shape of optimal trading strategies.
    
    We addressed the inverse problem, allowing traders to deduce the most likely adversary strategy given their own trading strategy. This approach provides a tool for traders to adapt their strategies in real-time as they gather more information about their competitors.

    We also discussed various approaches to strategy selection and then introduced risk-aversion into the equilibrium equations. 
    
    Overall, this paper contributes to the understanding of competitive trading by offering a robust theoretical foundation for strategy optimization in the presence of market impact, and provides practical insights into how traders can respond to competitive pressures in financial markets.

\section{Appendix: Mathematica code for two-trader equilibrium with risk}
\label{sec:appendix}

We used Wolfram Mathematica \cite{Mathematica} to solve \crefrange{eq:ddota-two-trader-with-risk}{eq:ddotb-two-trader-with-risk} but did not provide the explicit solutions. This is because they are fairly complex and their specific forms do not readily provide insight into the nature of the solutions. For completeness we provide the code used to generate Figure \ref{fig:equi-with-risk-1}. The other plots in Section \ref{sec:two-trader-equi-with-risk-aversion} were produced similarly.

\vskip 10pt

\begin{lstlisting}[language=Mathematica, mathescape=true]

(* Define constants *)
$\lambda$ = 5;
$\kappa$ = 5;
$\sigma$ = 0.5; (* Increased value for effect *)

(* Define the system of differential equations *)
solveSystem[$\xi_a$_, $\xi_b$_] := Module[{eqns, bcs, sol, aSol, bSol},
  eqns = {
    a$^{\prime\prime}$[t] == -($\lambda$/2) (b$^{\prime\prime}$[t] + $\kappa$ b$^\prime$[t]) + $\xi_a$ $\sigma^2$ a[t],
    b$^{\prime\prime}$[t] == -(1/(2 $\lambda$)) (a$^{\prime\prime}$[t] + $\kappa$ a$^\prime$[t]) + ($\xi_b$/$\lambda^2$) $\sigma^2$ b[t]
  };
  bcs = {a[0] == 0, a[1] == 1, b[0] == 0, b[1] == 1};
  sol = DSolve[{eqns, bcs}, {a[t], b[t]}, t];
  {a[t] /. sol[[1]], b[t] /. sol[[1]]}
];

(* Define the grid of $\xi_a$ and $\xi_b$ values, multiplied by 3 *)
xiValues = {
  {{1.5, 1.5}, {10, 10}, {50, 50}},
  {{10, 1.5}, {50, 1.5}, {200, 1.5}},
  {{1.5, 10}, {1.5, 50}, {1.5, 200}}
};

(* Generate the 3x3 grid of plots *)
gridPlots = Grid[Table[
   Module[{aSol, bSol, $\xi_a$ = xiValues[[i, j, 1]], $\xi_b$ = xiValues[[i, j, 2]]},
    {aSol, bSol} = solveSystem[$\xi_a$, $\xi_b$];
    Plot[{aSol, bSol}, {t, 0, 1}, 
     PlotLegends -> {"a(t)", "b(t)"}, 
     PlotStyle -> {Blue, Red}, 
     Frame -> True, 
     FrameLabel -> {{"a(t), b(t)", "$\kappa$=" <> ToString[$\kappa$] <>
       ", $\lambda$=" <> ToString[$\lambda$]}, {"t", "\(\xi_a=\)" <> ToString[$\xi_a$] <> 
       ", \(\xi_b=\)" <> ToString[$\xi_b$] <> ", $\sigma$=" <> ToString[$\sigma$]}}, 
     PlotRange -> {{0, 1},{0, 2.5}}]
    ], {i, 1, 3}, {j, 1, 3}], Frame -> All
];

(* Display the grid of plots *)
gridPlots
\end{lstlisting}

\section{Acknowledgements}

    I extend my sincere thanks to the Machine Learning Research Group at Morgan Stanley, to whom I gave two seminars on this work in August of 2024 during which they made many insightful suggests which greatly improved this work. I also extend my gratitude to Mike Shelley of the Courant Institute and Flatiron Institute for his invaluable guidance and Jim Gatheral of Baruch college for many illuminating discussions on transaction costs.

\bibliographystyle{alpha}
\bibliography{references}

\end{document}